\documentclass[twocolumn,amsmath,amssymb,aps,prd,groupedaddress]{revtex4-1}
\usepackage{graphicx}
\usepackage{graphics}
\usepackage[english]{babel}
\usepackage{dcolumn}
\usepackage{bm}
\usepackage{color}
\usepackage{soul}
\usepackage{textcomp}
 \usepackage{multirow}
\usepackage{soul}
\usepackage{rotating}
\usepackage{setspace}
\usepackage[mathlines]{lineno}
\usepackage{enumerate}
\usepackage{float}
\usepackage{blindtext}
\usepackage{subfigure}
\usepackage{microtype}
\usepackage[breaklinks=true,hyperindex=true,
pdftitle={Electromagnetically-induced frame dragging around astrophysical objects},
colorlinks=true,pagebackref=false,citecolor=blue,plainpages=false,pdfpagelabels,
linkcolor=blue,urlcolor=blue]{hyperref}

\def\mnras{Mon.~Not.~Roy.~Astr.~Soc.}
\def\apj{Astrophys. J.}
\def\apjl{Astrophys. J. Lett.}

\def\aap{Astron. Astrophys.}

\def\prd{Phys. Rev. D}

\begin{document}

\def\Xint#1{\mathchoice
{\XXint\displaystyle\textstyle{#1}}%
{\XXint\textstyle\scriptstyle{#1}}%
{\XXint\scriptstyle\scriptscriptstyle{#1}}%
{\XXint\scriptscriptstyle\scriptscriptstyle{#1}}%
\!\int}
\def\XXint#1#2#3{{\setbox0=\hbox{$#1{#2#3}{\int}$}
\vcenter{\hbox{$#2#3$}}\kern-.5\wd0}}
\def\ddashint{\Xint=}
\def\dashint{\Xint-}

\title{Electromagnetically-Induced Frame-Dragging around Astrophysical Objects}

\author{Andr\'es F. Guti\'errez-Ruiz}
\email{afelipe.gutierrez@udea.edu.co}
\affiliation{Grupo de F\'isica At\'omica y Molecular, Instituto de F\'{\i}sica,  
Facultad de Ciencias Exactas y Naturales, 
Universidad de Antioquia UdeA; Calle 70 No. 52-21, Medell\'in, Colombia.}

\author{Leonardo~A.~Pach\'on}
\email{leonardo.pachon@udea.edu.co}
\affiliation{Grupo de F\'isica At\'omica y Molecular, Instituto de F\'{\i}sica,  
Facultad de Ciencias Exactas y Naturales, 
Universidad de Antioquia UdeA; Calle 70 No. 52-21, Medell\'in, Colombia.}

\date{\today}

\begin{abstract}
Frame dragging (Lense-Thirring effect) is generally associated with 
rotating astrophysical objects.  
However, it can also be generated by electromagnetic fields if electric 
and magnetic fields are simultaneously present.
In most models of astrophysical objects, macroscopic charge neutrality 
is assumed and the entire electromagnetic field is characterized in 
terms of a magnetic dipole component. 
Hence, the purely electromagnetic contribution to the frame dragging
vanishes.
However, strange stars may possess independent electric dipole and 
neutron stars independent electric quadrupole moments that may lead 
to the presence of purely electromagnetic contributions to the frame 
dragging.
Moreover, recent observations have shown that in stars with strong 
electromagnetic fields, the magnetic quadrupole may have a significant 
contribution to the dynamics of stellar processes.
As an attempt to characterize and quantify the effect of electromagnetic 
frame-dragging in these kind of astrophysical objects, an analytic 
solution to the Einstein-Maxwell equations is constructed here on the 
basis that the electromagnetic field is generated by the combination 
of arbitrary magnetic and electric dipoles plus arbitrary magnetic 
and electric quadrupole moments.
The effect of each multipole contribution on the vorticity scalar and 
the Poynting vector is described in detail.
Corrections on important quantities such the innermost stable circular 
orbit (ISCO) and the epicyclic frequencies are also considered.
\end{abstract}


\maketitle

\section{Introduction}
Frame dragging (Lense-Thirring effect) is the quintessential hallmark 
of general relativity and is the result of the spacetime vorticity.
It was detected by the Gravity Probe B \cite{ED&11} and traditionally, 
it has been associated with rotating stellar astrophysical objects and in other 
astrophysical contexts, such as Galactic models, the frame dragging 
has been associated to the presence of magnetogravitational monopoles
\cite{CP13,CP14}.
Surprisingly, if the astrophysical object does not rotate but possesses 
both electric and magnetic fields, the spacetime vorticity does not 
vanish \cite{Das79,Bon91,HG&06}.
In this case, frame dragging is of purely electromagnetic nature and 
it is associated with the existence of a nonvanishing electromagnetic 
Poynting vector around the source \cite{Bon91,HG&06,MR&06,HB12,Her14}.
 
In most of the early models of astrophysical objects, macroscopic charge 
neutrality is assumed \cite{BBS96} and the magnetic field is characterized 
in terms of a pure dipole component \cite{Ghosh78, Ghosh79,Camenzind90, 
Konigl}.
This idea was endorsed by the confirmation of many predicted features 
of a star with a dipole field using two- and three-dimensional 
magnetohydrodynamic numerical simulations of magnetospheric accretion 
\cite{Romanova02, Romanova03, Romanova04,Long}. 
Thus, under this configuration the \emph{purely} electromagnetic 
contribution to the vorticity tensor vanishes.
However, astrophysical objects such as strange stars may possess independent 
electric dipole (see Sec.~10.4 in Ref.~\cite{Bec09}) and neutron stars 
independent electric quadrupole moments (see below), which may lead to 
the presence of purely electromagnetic contributions to frame dragging.
A precise account of those contributions is not available because of 
the lack of an analytic exact solution with such a complex electromagnetic 
field configuration.

Moreover, recent observations have shown that in stars with strong 
electromagnetic fields, the magnetic quadrupole may have  significant 
contributions on the dynamics of stellar processes.
Therefore, it is well known by now that the actual configuration of 
the magnetic field of strongly magnetized stars may depart from the 
dipole configuration \cite{Guvernondip}.
As an attempt to describe the spacetime geometry surrounding these kind 
of astrophysical objects, an analytic solution to the Einstein-Maxwell 
field equations is constructed here on the basis that the electromagnetic 
field is generated by the combination of arbitrary magnetic and electric 
dipole plus arbitrary magnetic and electric quadrupole moments.
This analytic exact solution allows for analyzing, e.g., the effect of 
each multipole contribution on the vorticity scalar and the Poynting 
vector (see Sec.~\ref{Sec:FrmDrggPntgVec}).
Moreover, it is possible to predict corrections to important quantities 
such as the innermost stable circular orbit  and the epicyclic frequencies.

To motivate further the derivation of the model considered here, the 
observational evidence for the existence of nondipolar fields in a 
variety of astrophysical objects is discussed next.

\section{Observational Evidence of Non-dipolar Fields}
Measurements of magnetic fields of strongly magnetized stars, based on 
the Zeeman--Doppler imaging technique \cite{lrsp-2005-8}, have shown 
that for these kind of astrophysical objects the magnetic field has a 
complicated multipolar topology in the vicinity of the star 
\cite{Donati97, Donati99,Jardine02,Jardine06,Johnstone2013}. 
This feature certainly is of prime relevance in, e.g., the accretion-disk 
dynamics in binary systems because if the quadrupole component dominates, 
then the flow of matter into the star will certainly differ from the 
well-known dynamics induced by a pure dipole field \cite{Jardine06}.

Complex configurations of magnetic fields are also present in stars
such as the T-Tauri stars.
They are young stellar objects of low mass that present variations 
in their luminosity. 
An important subclass of this kind of stars are the so-called classical 
T-Tauri stars (cTTS) because they present accretion from the circumstellar 
disk \cite{TTauriacre}.
Recent evidence points out that in classical T-Tauri stars, the magnetic
field near the star is strongly non-dipolar \cite{Johnstone2013}.
In the particular case of V2129 Oph, there exits a dominant octupole, 
0.12~T, and a weak dipole component, 0.035~T, of the magnetic field 
\cite{DJ&07}. 
Understanding the circumstellar disk dynamics, under complex field 
topologies, could provide insight into the formation of planets 
and the evolution of the star itself.

The discussion above also applies to white dwarfs. 
The first observations indicated that only a small fraction of white 
dwarfs appear to exhibit magnetic fields.
However, the observational situation changed significantly by the 
discovery of strong-field magnetic white dwarfs 
\cite{Schmidt03,1538-3881-130-2-734,0067-0049-204-1-5}, which are known 
to cover a wide range of field strengths $\sim1-100$T and deviates 
from the simple dipolar configuration \cite{refId02,refId03}.
Recent spectropolarimetric observations in white dwarf have shown
that, in addition to the dipole term, the quadrupole and octupole
terms make significant contributions to the field when it is
represented as an axisymmetric multipolar expansion \cite{Euchener05}.
Many studied cases indicate that higher multipole components or 
non axisymmetric components may be required in a
realistic model of white dwarfs (see, e.g., Ref \cite{Landstreet2000}).

Magnetars constitute an additional source of motivation.
They are characterized by their extremely powerful magnetic fields, 
covering strengths from $\sim10^8$ to $\sim10^{11}$~T 
\cite{McGillcatal,magnetars}, so that they can have occasional violent 
bursts. 
However, certain magnestars, such as the SGR~0418$+$5729, undergo this 
bursting phenomenon even with a weak magnetic fields ($\sim10^{8}$~T) 
\cite{Nrea2013}.
For SGR~0418$+$5729, the observed X-ray spectra cannot be fit 
with this low field strength; hence, it has been suggested that a 
hidden non-dipole field component must be present to explain the 
bursting episodes \cite{Guver21122011}.

In summary, there is sufficient observational evidence to develop 
a consistent analytic closed representation of the exterior 
spacetime around stars with non-dipolar magnetic fields.

\section[]{Analytic Formulae of the Model}
By combining the facts that (i) almost all the analytic closed form 
models for relevant astrophysical objects have been conceived in the 
frame of stationary axisymmetry geometry (see \cite{GPV13,PRV12,PRS06,
BG&09,JP11,TQ14} for the case of neutron stars), (ii) powerful tools 
to construct exact solution to the Einstein-Maxwell field equation 
have been developed, e.g. \cite{Ernst68b, Sib91,MS93}, and (iii) 
systematic studies on the construction of exact solution from its 
physical content have been performed \cite{PS06,Ernst06}, a new analytic 
exact solution to the Einstein-Maxwell field equations is introduced 
below.
This solution provides physical insight, e.g., into the influence of high 
order electromagnetic multipole moments in the frame dragging and 
in studying quasi-periodic oscillations (QPOs), which become a useful 
tool to identify the characteristics of the compact objects present in 
Low Mass X-Ray Binaries (LXRB) \cite{Stella99,1999ApJ...513..827M,
1998ApJ...492L..59S}. 
The model presented here is a member of the $N$-solitonic solution
derived in Ref.~\cite{RMM95}. 
In Appendix~\ref{app:metricfuncs} the relevant equations of the derived 
metric are summarized. 

In terms of the quasicylindrical Weyl-Lewis-Papapetrou coordinates 
$x^{\mu} = (t,\rho,z,\phi)$, the simplest form of the line element 
for the stationary axisymmetric case was given by Papapetrou 
\cite{Papap},
\begin{equation}
\label{Papapetrou} 
\mathrm{d}s^2= g_{\mu\nu} \mathrm{d} x^{\mu} \mathrm{d} x^{\nu},
\end{equation}
with $g_{tt} = - f(\rho,z)$, $g_{t\phi} = f(\rho,z) \omega(\rho,z)$,
$g_{\phi\phi} = \rho^2 f^{-1}(\rho,z) - f(\rho,z) \omega^2(\rho,z)$ and
$g_{zz}=g_{\rho\rho} = \mathrm{e}^{2\gamma(\rho,z)} f^{-1}(\rho,z)$. 
The metric functions $f$, $\omega$ and $\gamma$ can be obtained from 
the Ernst complex potentials ${\cal E}$ and $\Phi$ (see details in 
Ref.~\cite{Ernst68b}).
The Ernst potentials obey the relations \cite{Ernst68b}
\begin{equation}
\begin{split}
({\rm Re}\,{\cal E}+|\Phi|^2)\nabla^2{\cal E}&= (\nabla {\cal E} +
2\Phi^*\nabla\Phi)\cdot\nabla{\cal E}, 
\\ ({\rm Re}\,{\cal
E}+|\Phi|^2)\nabla^2\Phi&= (\nabla {\cal E} +
2\Phi^*\nabla\Phi)\cdot\nabla\Phi\, . 
\end{split}
\label{Ernst}
\end{equation}
From a physical viewpoint, the Ernst potentials are relevant because 
they lead to the definition of the analogues of the Newtonian gravitational 
potential, $\xi = (1- {\cal E} )/(1+{\cal E} )$,  and the Coulomb 
potential, $q = 2\Phi/ (1+{\cal E})$.
The real part of $\xi$ accounts for the matter distribution and its 
imaginary part for the mass currents. 
Besides, the real part of the $q$ potential denotes the electric field 
and its imaginary part the magnetic field.

The Ernst equations (\ref{Ernst}) can be solved by means of the 
Sibgatullin's integral method \cite{Sib91,MS93}, according to which 
the complex potentials $\cal E$ and $\Phi$ can be calculated from specified 
axis data ${\cal E}(z,\rho=0)$ and $\Phi(z,\rho=0)$ \cite{Sib91,MS93}.
Motivated by the accuracy \cite{PRS06,PRV12} and the level of generality 
of the analytic solution derived in Ref.~\cite{PRS06}, 
the Ernst potential ${\cal E}(z,\rho=0)$ is chosen as in Ref.~\cite{PRS06}.
To construct the exact solution that represents the electromagnetic 
field configuration described above, the Ernst potential $\Phi(z,\rho=0)$ 
is chosen following the prescription in Ref.~\cite{PS06}.
The Ernst potentials on the symmetry axis read
\begin{equation}
\begin{split}
{\cal E}(z,\rho=0) = \frac{z^3-z^2(m+ia)-kz+is}{z^3+z^2(m-ia)-kz+is}\, ,
\\
\Phi(z,\rho=0) = \frac{ z^{2}\varsigma+z(\upsilon + i \mu )+i \zeta +\chi}{z^3+z^2(m-ia)-kz+is}\, .
\end{split}
\label{Potencialeseje}
\end{equation}
The physical meaning of the parameters in Eq.~(\ref{Potencialeseje}) 
is derived from the multipole moments calculated using the 
Fodor-Hoenselaers-Perj\'es procedure \cite{Hoenselaers90} (see also 
Ref.~\cite{ST03}). 
For the present case, 
\begin{widetext}
\begin{equation}
\label{multipolosP}
\begin{split}
P_0 &= m , \qquad P_1=i a m,\qquad P_2 =
    (k-a^2)m,  \qquad P_3=-i m(a^3 - 2 a k + s),
\\
P_4&=\frac{1}{70} m \left[70 a^4-210 a^2 k+13 a \varsigma (\mu -i \upsilon )+140 a s 
 +10 k (7 k-m^2+\varsigma^2)+3 (\mu ^2+\upsilon ^2)\right],
  \\
P_5&=-\frac{1}{21} i m \left\{-21 a^5+84 a^3 k+a^2 (-6 \mu  \varsigma-63 s 
 +6 i \upsilon  \varsigma) \right.
\\
 &+a \left. \left[-63 k^2+6 k (m-\varsigma) (m+\varsigma)+\mu ^2 
 +5 i \zeta  \varsigma+\upsilon ^2+5 \chi  \varsigma\right] \right.
\\
 &+k \left. (3 \mu  \varsigma
 +42 s-i \upsilon  \varsigma)+2 i (\mu  \zeta +\upsilon  \chi )+7 s (\varsigma^2-m^2)
 \right\},
 \end{split}
\end{equation}
\begin{equation}
\label{multipolosQ}
\begin{split}
Q_0 &= \varsigma, 
\qquad 
Q_1 =\upsilon+i(a \varsigma+\mu), 
\qquad  
Q_2 =-a^2 \varsigma-a \mu +k \varsigma+\chi+i(a \upsilon +\zeta),
\\ 
Q_3 &=-a^2 \upsilon -a \zeta +k \upsilon+i(-a^3 \varsigma-a^2 \mu +
a (2 k \varsigma+\chi )+k \mu -s \varsigma),
\\
Q_4 &= a^4 \varsigma+a^3 (\mu -i \upsilon )+a^2 (-3 k \varsigma-i \zeta -\chi )
+\frac{1}{70} a 
\left[ 140 s \varsigma-(\mu -i \upsilon ) (140 k-3 m^2-10 \varsigma^2)\right]
\\ 
&+\frac{1}{7} 
\left\{
\varsigma \left[\varsigma (k \varsigma+i \zeta +\chi ) 
+(\mu -i \upsilon )^2\right]+(7 k-m^2) (k \varsigma+i \zeta +\chi )+7 s (\mu -i \upsilon )
\right\},
\\
Q_5&=\frac{1}{21} 
\left\{
  21 i a^5 \varsigma+21 a^4 (\upsilon +i \mu ) -21 i a^3 (4 k \varsigma+i \zeta +\chi ) 
  +a^2 \left[ 63 i s \varsigma-i (63 k+8 \varsigma^2) (\mu -i \upsilon )\right]
\right.
\\ 
    &+ \left. i a \left\{-\varsigma \left[ \varsigma (8 k \varsigma+i \zeta +\chi ) 
    +9 \mu ^2-16 i \mu  \upsilon -7 \upsilon ^2 \right] 
    +(21 k-2 m^2) (3 k \varsigma+2 i \zeta +2 \chi ) 
    \right.
\right.    
\\ 
    &+\left. \left. 42 s (\mu -i \upsilon )\right\}
    -i (\mu -i \upsilon ) (-21 k^2+2 k m^2+\mu ^2+\upsilon ^2)
    +k \varsigma^2 (\upsilon -i \mu )
\right.    
\\ 
    &+\left. \varsigma (-42 i k s-6 \mu  \zeta +6 i \mu  \chi +7 i m^2 s
    + 8 i \zeta  \upsilon +8 \upsilon  \chi )+21 s (\zeta -i \chi )+7 i s \varsigma^3 
\right\}.
\end{split}
\end{equation}
\end{widetext}
Specifically, the interpretation of the parameters based on the multipole 
expansion in Eqs.~(\ref{multipolosP}) and (\ref{multipolosQ}) is as follows.
The real parameter $m$ corresponds to the total mass, $a$ to the total 
angular moment per unit mass while $k$ and $s$ are related to the 
mass-quadrupole moment and the differential rotations, respectively. 
For later convenience, an electric monopole contribution $Q_0$, 
characterized by the parameter $\varsigma$, was introduced above.
Parameters $\upsilon$ and $\mu$ are associated with the electric and 
magnetic dipole moments, respectively, whereas $\chi$ and $\zeta$ 
with the electric and magnetic quadrupole moments, respectively.
The existence of an electric dipole is theorized for strange stars (See 
Ref.~\cite{Wernerbook}).
A summary of the arbitrary parameters and the multipole moments 
they are related to can be found in Table~\ref{tmult}.
\begin{table}
\begin{tabular}{|c|c|}
\hline \hline
 Symbol       & Associated Multipole Moment
\\ \hline \hline
$\varsigma$   & Electric monopole   \\   
$\upsilon$        & Electric dipole          \\
$\chi$        & Electric quadrupole   \\ 
$\mu$         & Magnetic dipole     \\
$\zeta$       & Magnetic quadrupole  \\ \hline\hline
\end{tabular}
\caption{Summary of the electromagnetic parameters and the 
multipole moments they are related to.}
\label{tmult}
\end{table}

The mass moment $P_2$ governs the deformation of the star and it is 
composed of two parts: the term $a^{2}m$ that is the usual 
rotation-induced deformation and a second contribution $km$ that accounts 
for a possible intrinsic deformation of the star \cite{Bocquet95,2007PhRvD..75b3008D}.
An analogous argument can be formulated in the case of the electric 
moments. 
The real part of $Q_2$ accounts for the electric quadrupole contribution to 
the total electromagnetic quadrupole moment.
The terms $-a^2 \varsigma$ and $-a\mu$ account for the rotation-induced 
redistribution of the electric charge and deformation of the magnetic dipole; 
whereas, the term $k \varsigma$ accounts for the contribution from 
the charge distributed over the intrinsic deformed mass.  
The additional parameter $\chi$ is added to account for any additional possible 
contribution to the total quadrupole moment.

The multipole expansion in Eq.~(\ref{multipolosQ}) shows that even 
if the magnetic dipole parameter is zero ($\mu=0$), a magnetic dipole 
component (imaginary part of $Q_1$) is present provided by the rotation 
of the electric charge $Q_0$. 
Similarly, even if the magnetic quadropole parameter $\zeta$ is set 
to zero, a rotating electric dipole can induce a magnetic quadrupole 
(imaginary part of $Q_2$).
For the electric part of the multipole expansion in Eq.~(\ref{multipolosQ}), 
an analogous behavior is observed, namely, a rotating magnetic dipole
can induce an electric quadropole moment and a rotating magnetic
quadrupole can generate an electric octupole moment.
Based on these processes, the astrophysical source can afford a 
non-vanishing induced Poynting vector and, correspondingly, an 
induced non-vanishing flux of electromagnetic energy around the 
source that will contribute to the frame dragging induced solely by 
the mass currents.

\section{Characterization of the Electromagnetic Fields}
\label{sec:Properties}
To describe the electromagnetic properties of the solution, the electric 
and magnetic fields, in the spacetime surrounding the star, are 
calculated by means of the expressions
\begin{equation}\label{EB}
E_{\alpha} = F_{\alpha \beta}u^{\beta}\, ,\qquad B_{\alpha} =
-\frac{1}{2}\,\epsilon_{\alpha \beta}^{\hphantom{\alpha \beta}\gamma
\delta} F_{\gamma \delta}u^{\beta}\, ,
\end{equation}
where $F_{\alpha \beta}$ is the electromagnetic field tensor
$F_{\alpha \beta} = 2\,A_{[\beta;\alpha]}$,
$A_{\mu}=(0,0,A_{\phi},-A_t)$ is the electromagnetic four-potential,
$u_{\alpha}$ is a time--like vector and $\epsilon_{\alpha 
\beta \gamma \delta}$ is the totally antisymmetric tensor of positive 
orientation with norm $\epsilon_{\alpha \beta \gamma \delta}
\epsilon^{\alpha \beta \gamma \delta} = -24$ \cite{Wal84}.
For a congruence of observers at rest in the frame of (\ref{Papapetrou}),
the four--velocity is defined by the time--like vector
$u^\alpha=(1/\sqrt{f},0,0,0)$.
The vectorial fields have components in the $\rho$ and $z$
directions only. 
The components of the electric field are given by
\begin{equation}\label{Eexplicito}
E_\rho = -\frac{\sqrt{f}}{e^{2 \gamma}}\,A_{t, \rho}\, ,\qquad E_z =
-\frac{\sqrt{f}}{e^{2 \gamma}}\,A_{t, z}\, ,
\end{equation}
and for the magnetic field by
\begin{eqnarray}\label{Bexplicito}
B_\rho &=& \frac{f^{3/2}}{\rho e^{2 \gamma}}\,(-\omega A_{t,z} +
A_{\phi,z})\, ,\\B_z &=& -\frac{f^{3/2}}{\rho e^{2
\gamma}}\,(-\omega A_{t,\rho} + A_{\phi,\rho})\, .
\end{eqnarray}
The explicit form of the fields can be found in Appendix~\ref{app:emfield}.

\begin{center}
\begin{figure}[h!]
\begin{center}
\begin{tabular}{cc}
\includegraphics[width=0.5\columnwidth]{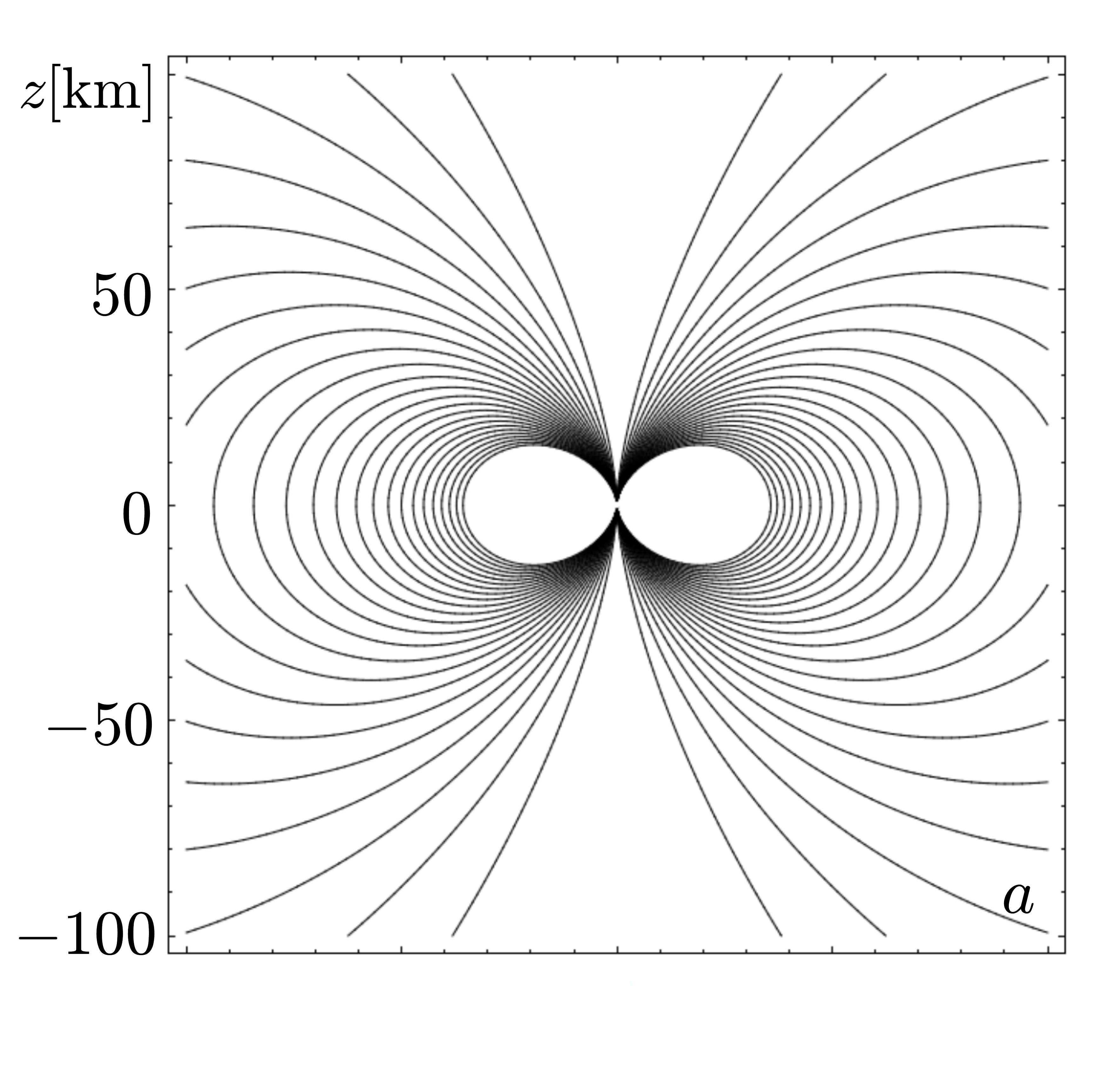}
&
\hspace{-0.025cm}\includegraphics[width=0.5\columnwidth]{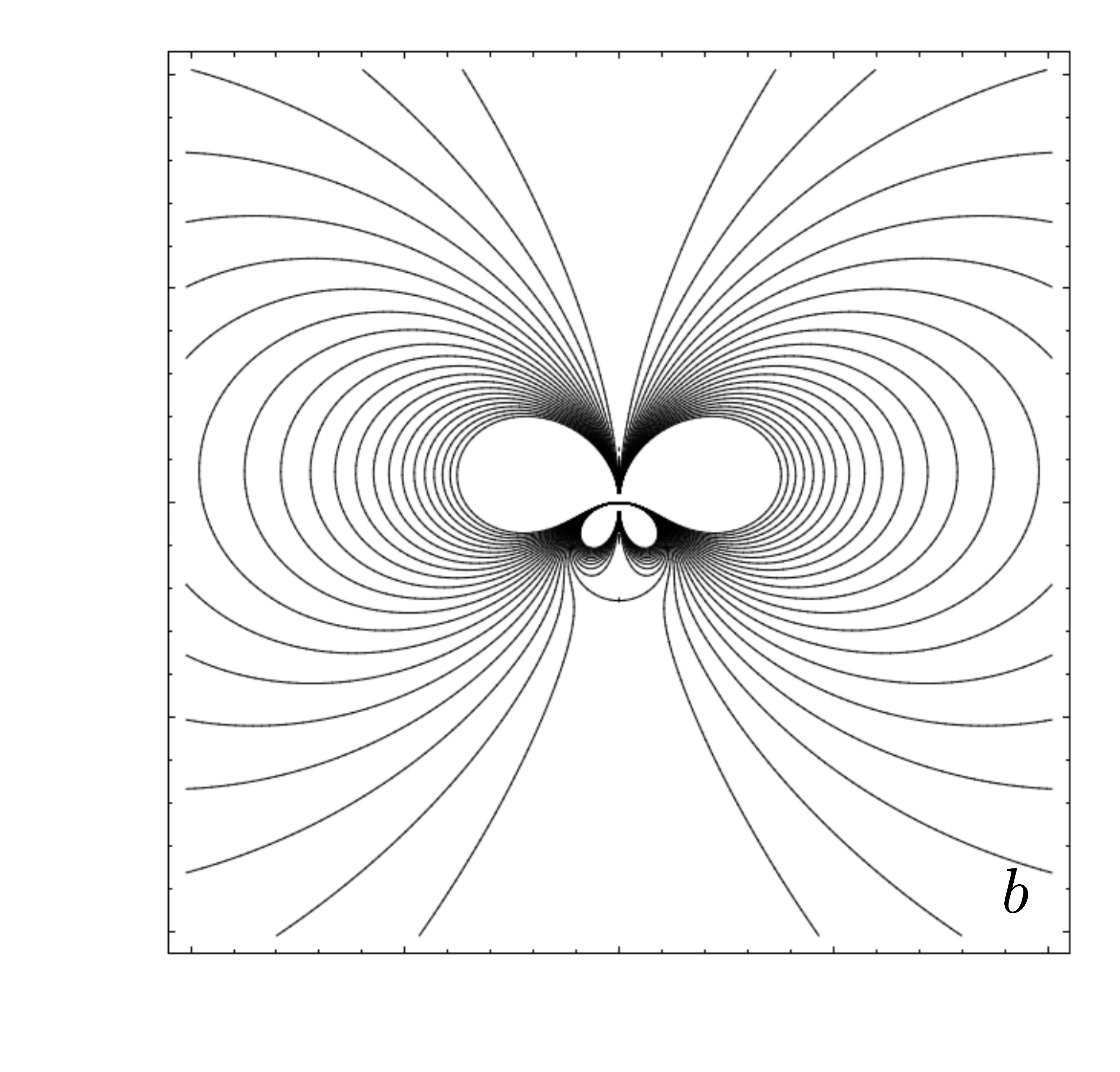}
\vspace{-0.5cm} 
\\ 
\includegraphics[width=0.5\columnwidth]{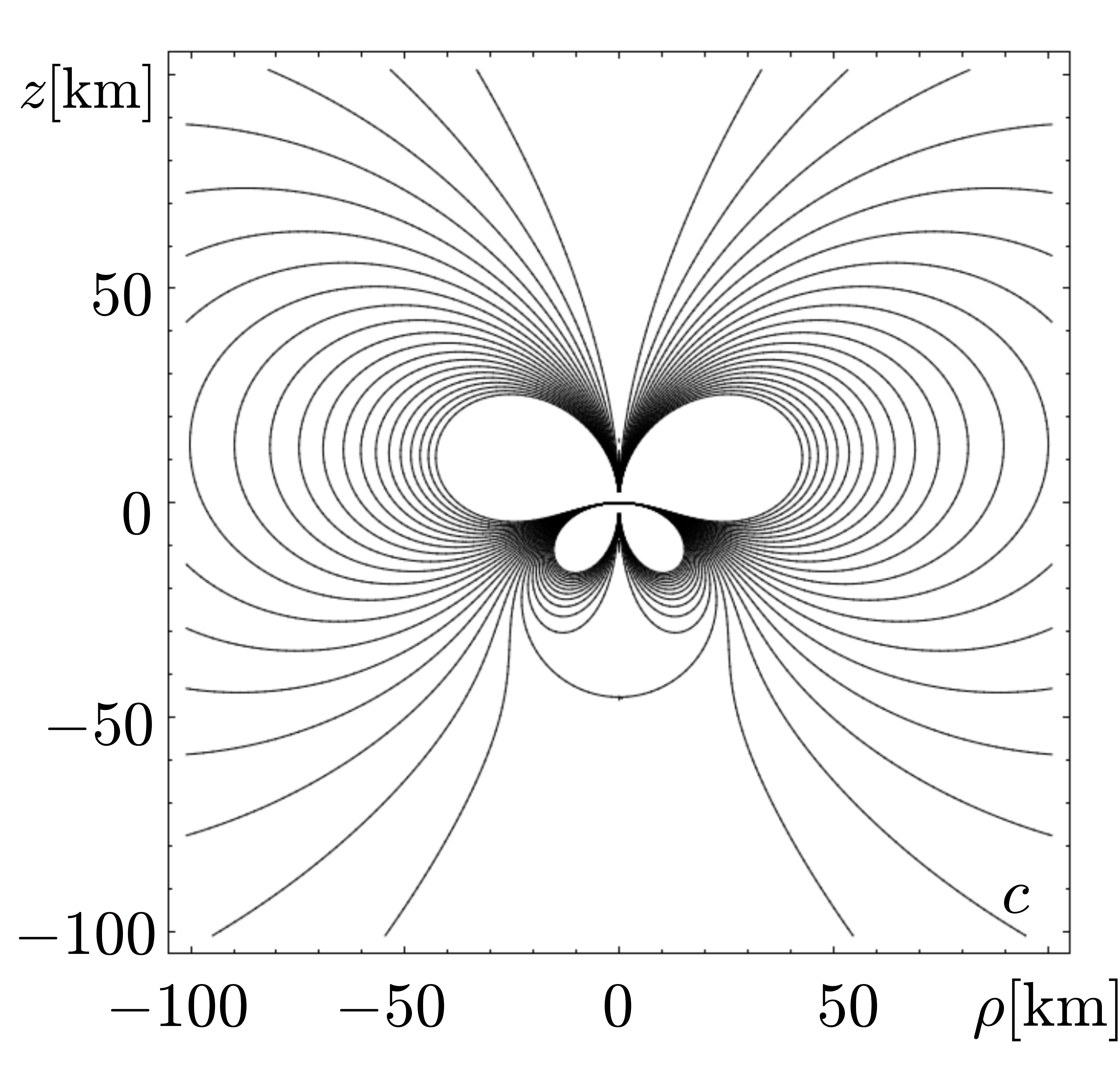}
& 
\hspace{-0.25cm} \includegraphics[width=0.5\columnwidth]{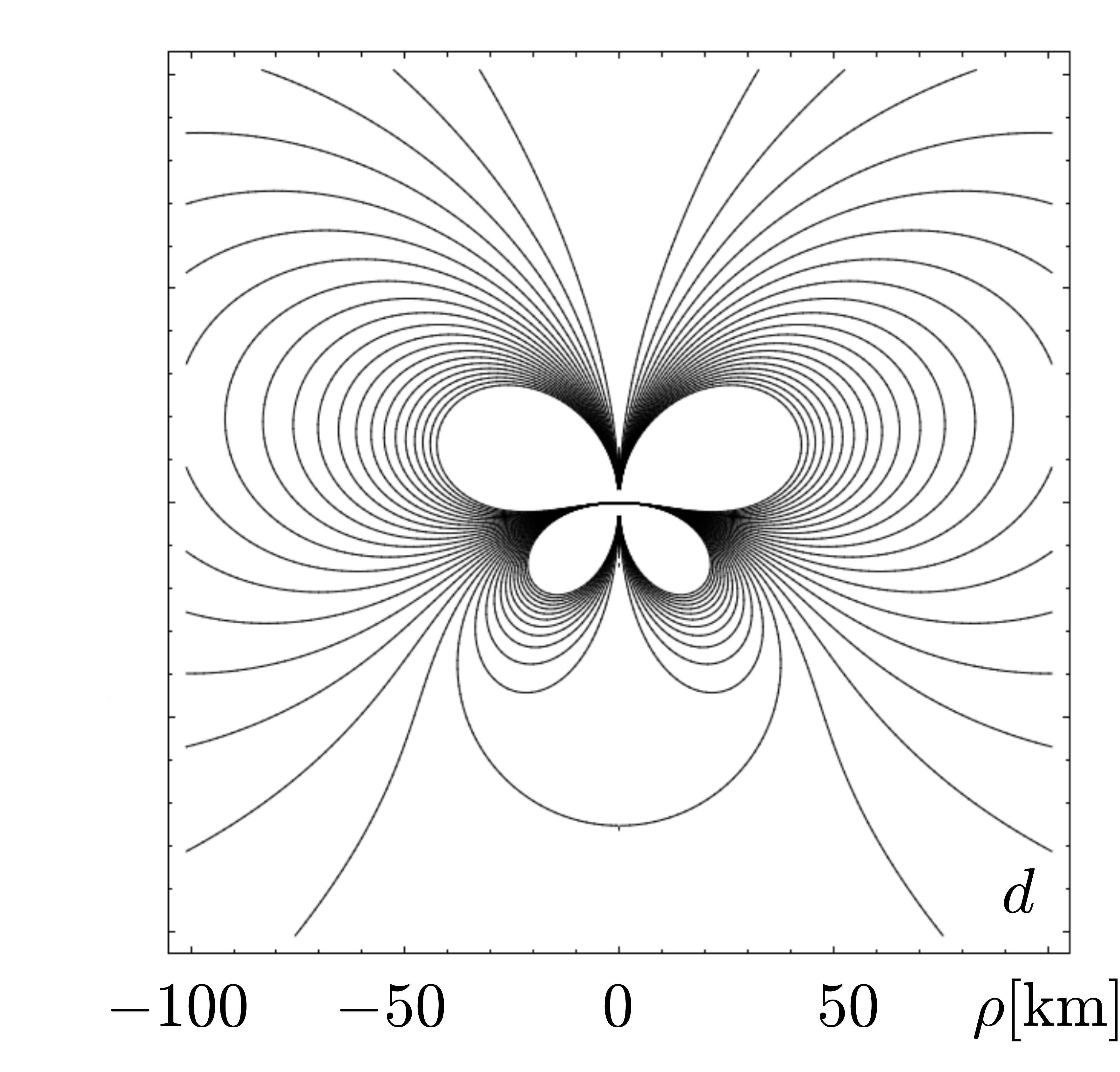}
\end{tabular}
\end{center}
\caption{
Magnetic field force lines for
$m =2.071$~km, $j = 0.194$, $Q = -2.76$~km$^3$, $s = -2.28$~km$^4$, 
$\varsigma = 0$~km, $\upsilon=0$~km$^2$, $\chi=0$, $\mu=1$~km$^2$ 
for (a) $\zeta=0$~km$^3$, (b) $\zeta=15$~km$^3$, (c) $\zeta=30$~km$^3$ 
and (d) $\zeta=50$~km$^3$.
The non-electromagnetic parameters correspond to the model 2 for the equation 
of state L in Ref.~\cite{Pappas12a} (see also Table~\ref{tp}).
}
\label{fig:magneticquadrupole}
\end{figure}
\end{center}
Figure~\ref{fig:magneticquadrupole} shows the force lines of the magnetic 
field for various values of the magnetic quadrupole parameter $\zeta$ and 
for realistic values of the mass and mass current multipoles.
Specifically, the vacuum multipole moments of the solution mass, angular moment 
and mass quadrupole and current octupole have been fixed to the numerical 
ones obtained in Ref.~\cite{Pappas12a}. 
They are listed in Table~\ref{tp}.
\begin{table}
\begin{center}
\begin{tabular}{ccccc}\hline\hline
 Model       & $m$ [km] & $j$  & $Q$ [km$^3$]& $s$ [km$^4$] \\ \hline\hline
M2 &  2.071 & 0.194 &  -2.76  & -2.28 \\ 
M3 &  2.075 & 0.324 & -7.55   & -10.5 \\ 
M4 &  2.080 & 0.417 & -12.2   & -22.0 \\
M5 &  2.083 & 0.483 & -16.2   & -33.9 \\ \hline\hline
\end{tabular}
\end{center}
\caption{Realistic numerical solutions for rotating neutron stars 
derived by Pappas and Apostolatos~\cite{Pappas12a} for the equation 
of state L. 
Here, $m$ is the total mass of the star, $j$ is the dimensionless 
spin parameter: $j=J/M^2$ (being $J$ the angular momentum), $Q$ is 
the quadrupole moment and $s$ is the current octupole moment. 
See Table VI of \cite{Pappas12a}.}
\label{tp}
\end{table}
In particular, for Fig.~\ref{fig:magneticquadrupole}.a, 
$\zeta=0$~km$^3$; \ref{fig:magneticquadrupole}.b, $\zeta=10$~km$^3$; 
\ref{fig:magneticquadrupole}.c, $\zeta=25$~km$^3$ and for 
Fig.~\ref{fig:magneticquadrupole}.d, $\zeta=50$~km$^3$.
The increasing of the separation between consecutive force lines indicates
that the magnetic field  decreases while the distance increases. 
Figure~\ref{fig:magneticquadrupole} not only shows how the reflection 
symmetry around the plane $z=0$ is broken because of the magnetic 
quadrupole \cite{HG&06}, but also shows that at large distances from 
the source, the magnetic field behaves like a magnetic dipole despite 
the presence of strong non-dipolar contributions.

\begin{center}
\begin{figure}
\begin{center}
\begin{tabular}{cc}
\includegraphics[width=0.5\columnwidth]{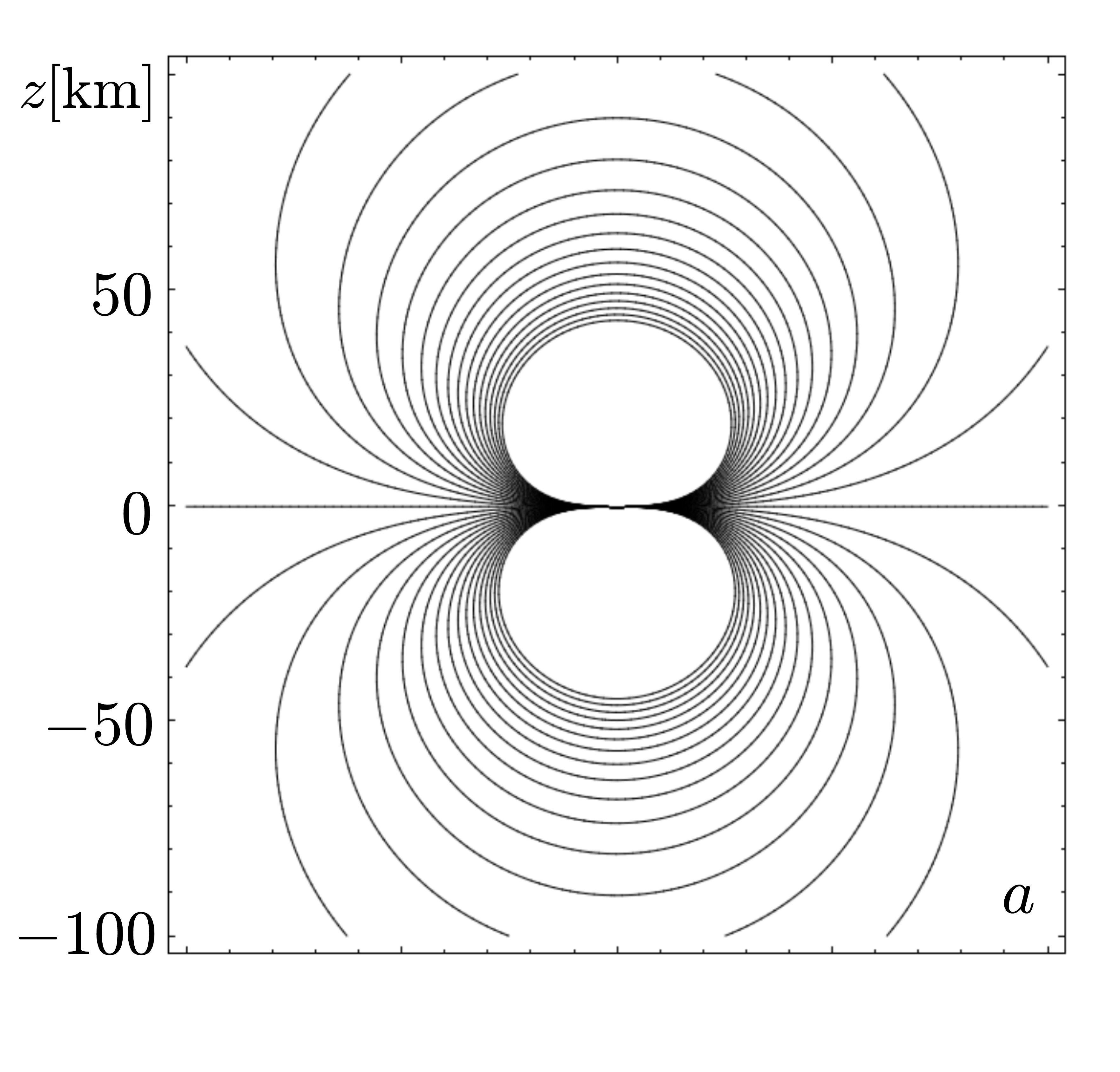}
&\hspace{-0.135cm} \includegraphics[width=0.5\columnwidth]{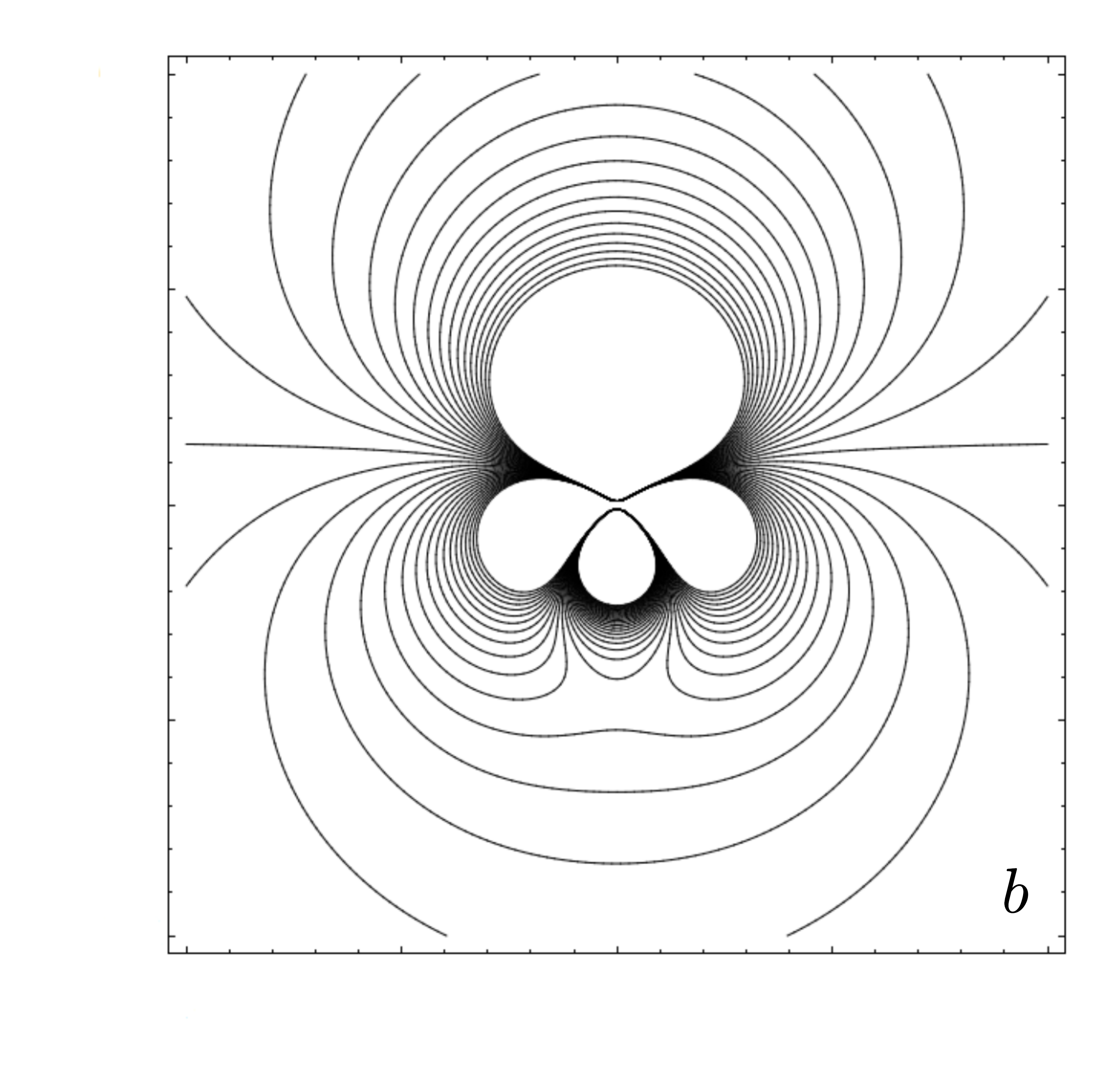}
\vspace{-0.5cm}  \\
\includegraphics[width=0.5\columnwidth]{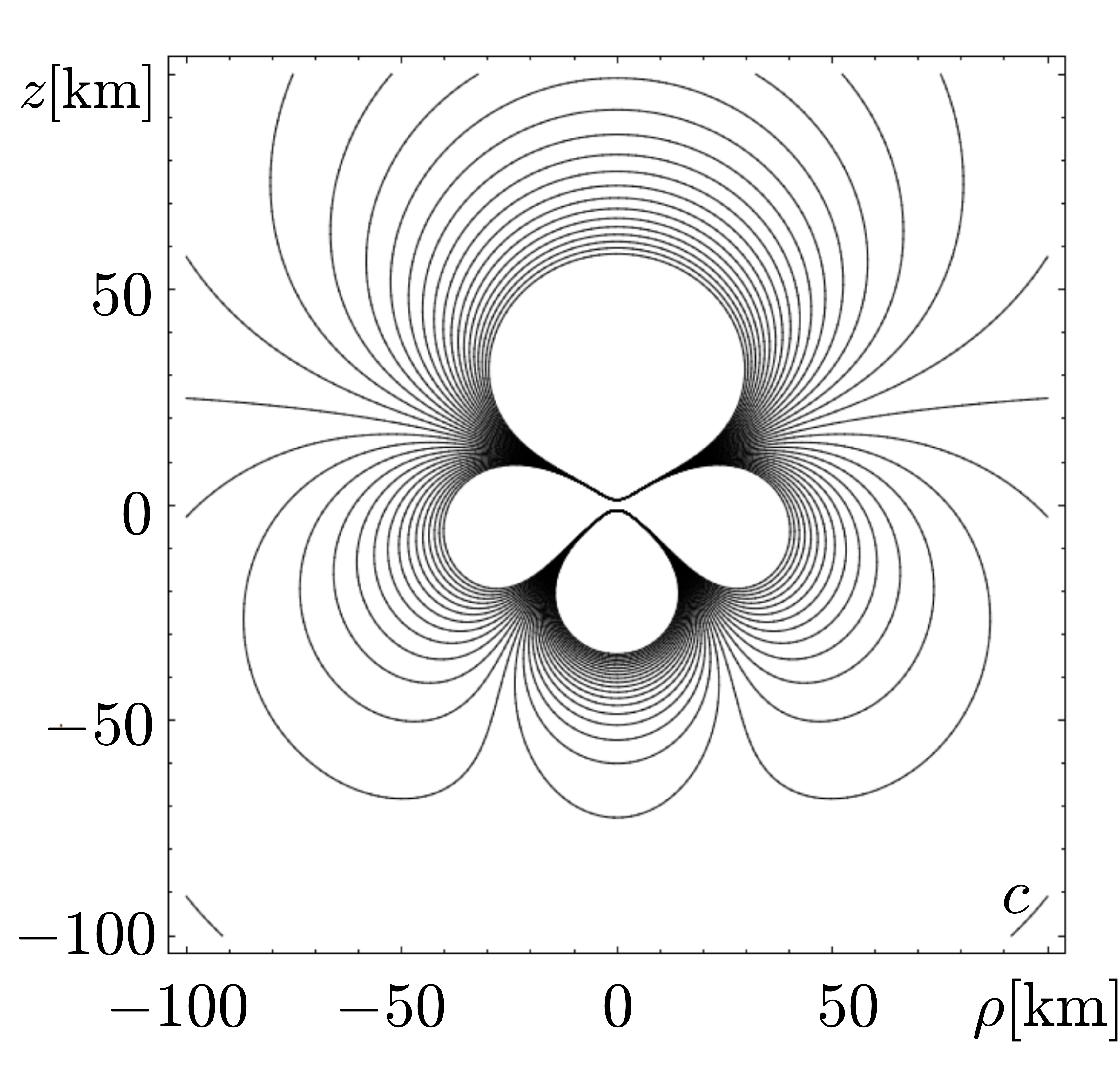}
& \hspace{-0.25cm} \includegraphics[width=0.5\columnwidth]{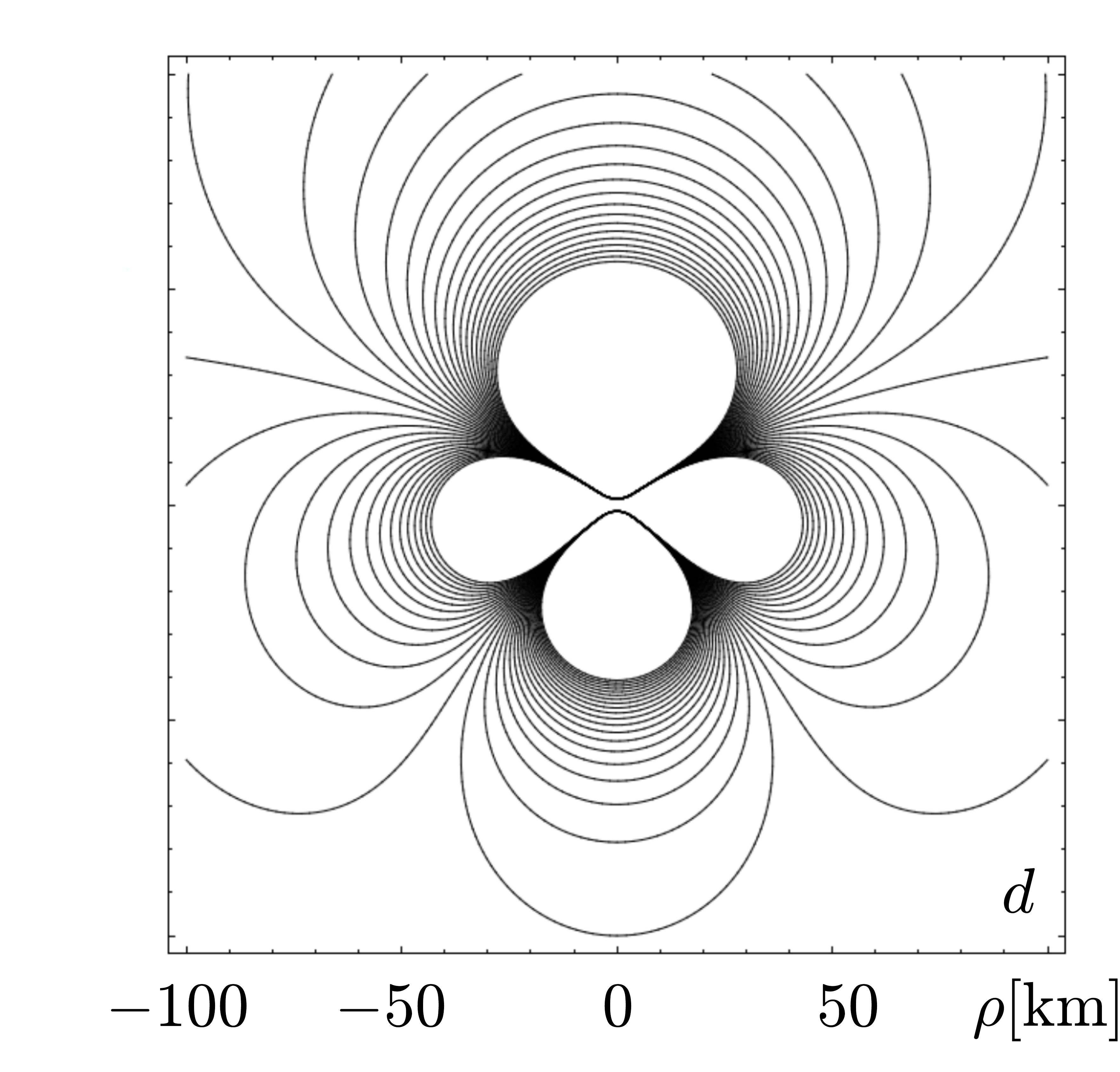}
\end{tabular}
\end{center}
\caption{
Electric field force lines for 
$m =2.071$~km, $j = 0.194$, $Q = -2.76$~km$^3$, $s = -2.28$~km$^4$, 
$\varsigma = 0$~km, $\upsilon=1$~km$^2$, $\mu=0$~km$^2$, $\zeta=0$ 
for (a) $\chi=0$~km$^3$, (b) $\chi=10$~km$^3$, (c) $\chi=25$~km$^3$ 
and (d) $\chi=50$~km$^3$.
}
\label{fig:electricquadrupole}
\end{figure}
\end{center}
Figure~\ref{fig:electricquadrupole} shows the force lines of the 
electric field for a variety of values of the electric quadrupole 
for realistic values of the mass and mass current multipoles
listed in Table~\ref{tp}.
Specifically, for Fig.~\ref{fig:electricquadrupole}.a, $\chi=0$~km$^3$; 
\ref{fig:electricquadrupole}.b, $\chi=10$~km$^3$; \ref{fig:electricquadrupole}.c, 
$\chi=25$~km$^3$ and for Fig.~\ref{fig:electricquadrupole}.d, 
$\chi=50$~km$^3$.
As above, it is clear that reflection symmetry is broken and that
at large distances the electric dipole component dominates the 
field configuration.

In general, the electromagnetic field of astrophysical objects is
expected to be a combination of the results depicted in 
Figs.~\ref{fig:magneticquadrupole} and  \ref{fig:electricquadrupole}.
Moreover, as shown below, the breaking of the reflection symmetry 
has an important role on the vorticity of the spacetime.
After characterize the electromagnetic fields, the generation 
of purely electromagnetic frame dragging is discussed below.

\section{Vorticity Scalar and Poynting Vector}
\label{Sec:FrmDrggPntgVec}
The physics of the frame dragging states that the rotation of the source 
induces a twist in the neighbourhood that drags  any frame  of reference  
near the source. 
In the case of spacetimes endowed by complex electromagnetic fields 
and mass currents, frame dragging originates from a combination of 
the vorticity of the electromagnetic field and the vorticity associated 
with the mass currents. 

\begin{center}
\begin{figure}[h!]
\begin{center}
\begin{tabular}{cc}
{\includegraphics[width=0.5125\columnwidth]{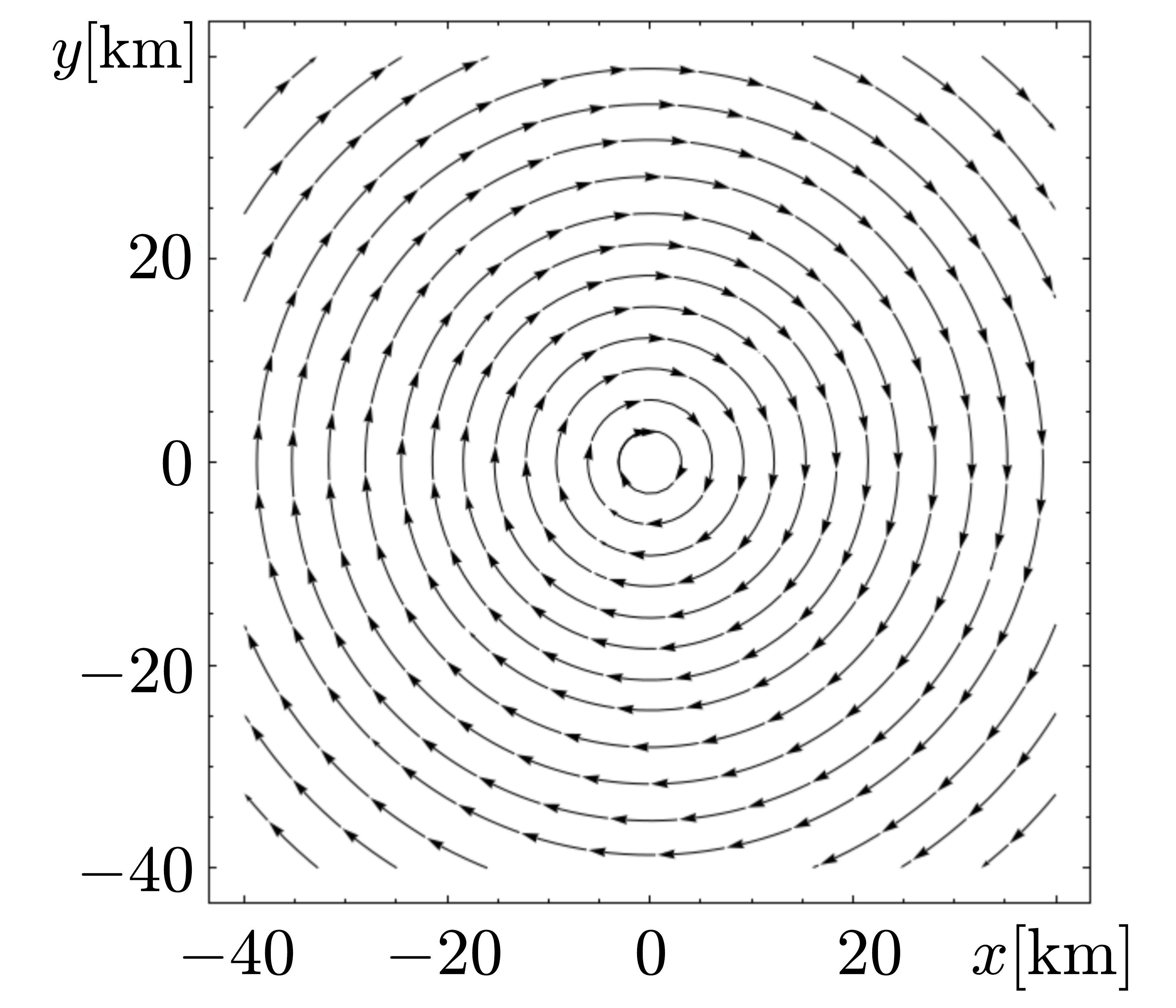}}
\hspace{-0.4cm}
&
{\includegraphics[width=0.5125\columnwidth]{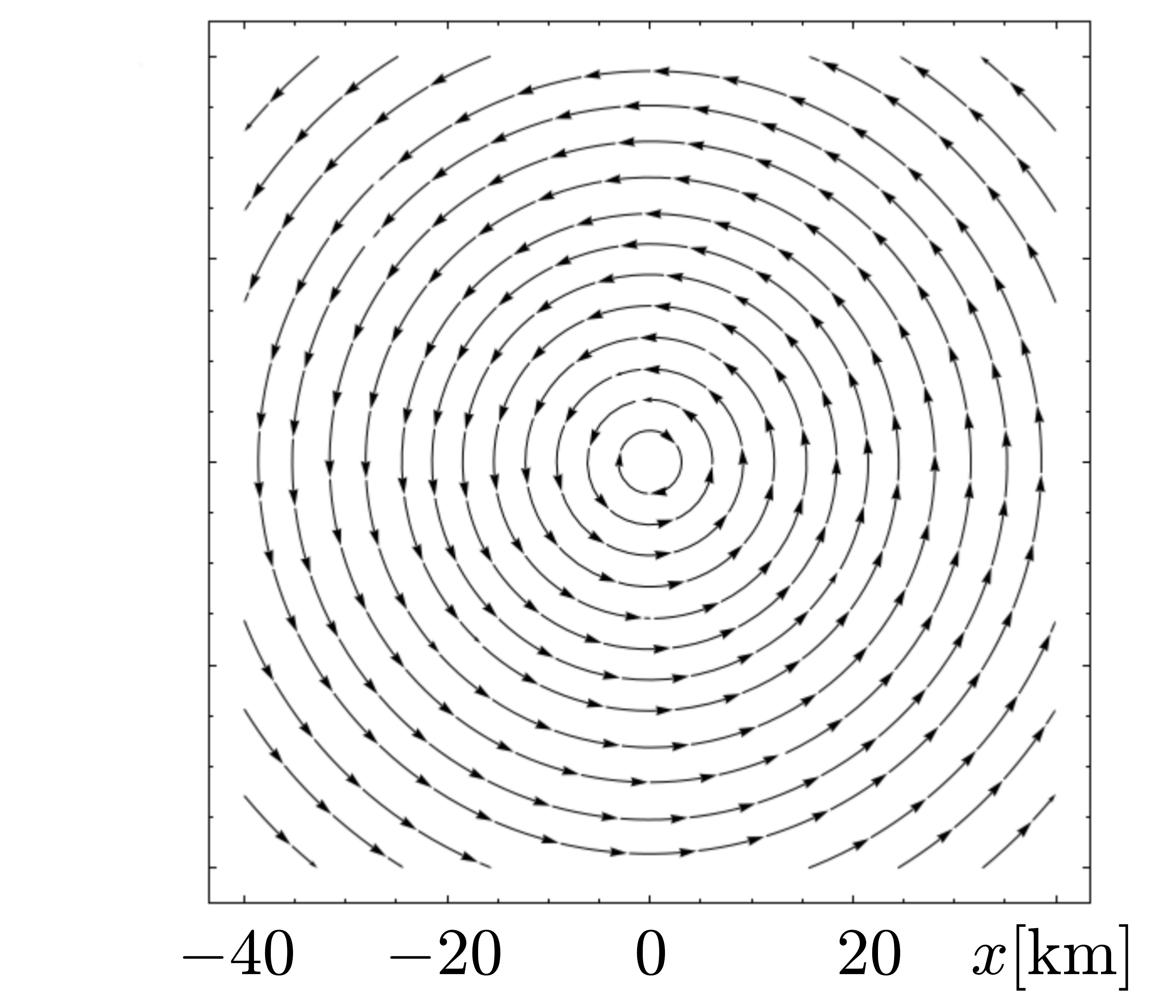}
}
\end{tabular}
\end{center}
\caption{
Vector field of the Poynting vector for the star model  $m =2.071 ~\mathrm{km}$, $j =
0.194$, $Q = -2.76 ~\mathrm{km}^3$, $s = -2.28 ~\mathrm{km}^4$.
The electromagnetic multipoles were chosen zero except
$\varsigma=0.1 ~\mathrm{km}$, $\mu=1 ~\mathrm{km}^2$ in the left panel and 
$\mu=-1 ~\mathrm{km}^2$ in the right side. 
The direction of the Poynting vector is due the sign change in the multipole 
expansions in Eq.~(\ref{multipolosQ}).
Figures are presented in the quasi-Cartesian auxiliary coordinates 
$r\rightarrow \sqrt{x^{2}+y^{2}}$ and $\phi=\mathrm{arctan}(y/x)$.
}
\label{fig:pyntngvctr}
\end{figure}
\end{center}

\emph{Poynting vector.}\textendash The Poynting vector $\mathbf{S}$ carries 
the information about the electromagnetic energy flux in the spacetime. 
Because of the axially symmetric character of  the spacetime, only 
the component  along the unitary vector $\hat{\mathrm{e}}_\phi$ survives
\cite{HG&06}.
In terms of the Ernst potentials 
$S^{\phi}=\sqrt{f}\mathrm{Im}\left(\Phi^{*}_{,\rho}\Phi_{,z}\right)
/(4\pi \rho e^{2\gamma})$ \cite{Manko06}
or more conveniently
\begin{align}
  S^{\phi}&=\frac{\sqrt{f}}{8\pi \rho e^{2\gamma}} \nabla \Phi^* \times \nabla \Phi,
 \label{equ:poynvec}
\end{align}
where it is clear that $S^{\phi}$ vanishes if $\Phi$ is purely real, 
purely imaginary or when their real and imaginary parts are proportional 
to each other. 
Due to the complex combination of the electromagnetic moments with the 
mass currents, their effects on the Poynting vector (and subsequently 
to the vorticity) are often subtle.
However, the multipole expansion in Eq.~(\ref{multipolosQ}) allows 
for a detailed description of each contribution.
For instance, if $\upsilon=0$, $\chi=0$ and $|a\varsigma| < |\mu|$, it is 
then clear that, to leading order, a sign change in the magnetic dipole 
moment parameter $\mu$ changes the sign of the  magnetic field.
This has the effect of changing the rotation direction of the Poynting 
vector (see Fig.~\ref{fig:pyntngvctr}) and as discussed below, it could
decrease the total vorticity of the spacetime.
Figure~(\ref{fig:pyntngvctr}) depicts the Poynting vector circulation 
around the source, when $\mu$ is chosen parallel (l.h.s. panel) and 
antiparallel (r.h.s. panel) to the star's rotation axis.

The change of the circulation of the Poynting vector can be undesrtood
from the potential $q(\rho,z)$ that is analoguos to the Coloumb potential
(see above). 
Under the condition that the reflection symmetry is not broken, $\upsilon=0$
and $\zeta=0$, a sign change of the magnetic dipole parameter $\mu$ is 
equivalent to change the sign of the coordiante $z$ with a global minus 
sign, namely, $q(\rho,z;\upsilon=0,\zeta=0,-\mu) = 
-q(\rho,-z;\upsilon=0,\zeta=0,\mu) = -q^*(\rho,z;\upsilon=0,\zeta=0,\mu)$.
Therefore, under the conditions that reflection symmetry imposses \cite{PS06},
all the electric field moments change sing and this leads to a global
sing of the Poynting vector.

\emph{Vorticity Scalar.}\textendash 
For an observer at rest respect to (\ref{Papapetrou}), the vorticity 
tensor is defined by
$
    \omega_{\alpha\beta} = u_{[\alpha;\beta]} + \dot{u}_{[\alpha}u_{\beta]}.
    \label{VorticityTensor}
$
The direct calculation of $ \omega_{\alpha\beta}$ for (\ref{Papapetrou}) 
yields to \cite{HG&06}
\begin{equation}
    \omega_{\alpha\beta} = \left(\begin{tabular}{cccc}
    0 & 0 & 0 & 0 \\
    0 & 0 & 0 & $-\frac{1}{2}\sqrt{f}\omega_{,\rho}$ \\
    0 & 0 & 0 & $-\frac{1}{2}\sqrt{f}\omega_{,z}$ \\
    0 & $\frac{1}{2}\sqrt{f}\omega_{,\rho}$ &
    $\frac{1}{2}\sqrt{f}\omega_{,z}$ & 0
\end{tabular}\right) .
\label{vorticitytensorgeneral}
\end{equation}
To quantify the vorticity, it is convenient to introduce the vorticity 
scalar that is defined by the contraction of the vorticity tensor and 
reads
\begin{equation}
 \omega_{\mathrm{v}}
 =
\left(\omega^{\alpha}_{\beta}\omega^{\beta}_{\alpha}\right)^{\frac{1}{2}}
 =f\sqrt{f(\omega_{,\rho}^{2}+\omega_{,z}^{2})}/\sqrt{2 e^{2\gamma}\rho^2} ,
 \label{vorscalar}
\end{equation}
with $\omega_{,\rho}=-\rho f^{-2} \Im({\cal E}_{,z} + 2 \Phi^{*}
    \Phi_{,z})$ and $\omega_{,z}= \rho f^{-2} \Im ({\cal E}_{,\rho} + 2 \Phi^{*}
    \Phi_{,\rho}).$
More explicitly, it reads
\begin{equation}
 \omega_{\mathrm{v}}=\frac{e^{-\gamma}}{\sqrt{2f}}\sqrt{
 \Im[\mathcal{E}_{,z}
 +2\Phi^{*}\Phi_{,z}]^{2}+\Im[\mathcal{E}_{,\rho}+2\Phi^{*}\Phi_{,\rho}]^{2} }
  \label{equ:vorticity}
\end{equation}
The vorticity scalar can be understood in terms of its fluid mechanics 
analogue. 
It represents the rotation of the fluid. 
In the general relativistic case, it can be related to the rotation velocity 
of a family of congruences.

Equation~(\ref{equ:vorticity}) is very useful to analyze the contributions 
to the vorticity because the electromagnetic and mass currents terms can be easily identified 
there. 
Moreover, this identification can be accompanied by further expressing
these contributions in terms of the parameters of the Ernst potentials~(\ref{Potencialeseje}).
In particular, in absence of electromagnetic fields, the imaginary part 
of $\mathcal{E}$ is associated with mass currents.
Thus, for static sources one could attempt to assume $\mathcal{E}$ as real 
in Eq.~(\ref{equ:vorticity}) and focus only in the electromagnetic 
contributions encoded in $\Phi$.
However, because $\mathcal{E}$ and $\Phi$ are not independent objects 
[see, e.g., Eq.~(\ref{Ernst})], in the presence of electromagnetic fields, 
$\mathcal{E}$ and $\Phi$ are complex even for non-spinning astrophysical 
objects.
Notwithstanding, if $\Phi$ is purely real or purely imaginary, for 
non-rotating objects, the imaginary part of $\mathcal{E}$ vanishes
[see, e.g., Eq.~(\ref{Ernst})].
Therefore, interest here is in identifying when the electromagnetic 
contribution does not vanish. 

In doing so, it is convenient to express 
$ \omega_{\mathrm{v}} = |\Im (\nabla \mathcal{E} 
+ 2\Phi^* \nabla \Phi) |/(\sqrt{2f}e^{\gamma})$, so that the term 
$\Im \Phi^* \nabla \Phi$ vanishes if $\Phi$ is purely real, 
purely imaginary or when their real and imaginary parts are proportional 
to each other.
Based on the discussion above, see also Eq.~(\ref{multipolosQ}), the 
purely real case corresponds to the absence of magnetic fields and 
the purely imaginary case to the absence of electric fields.
In the cases when the Poynting vector is zero, the electromagnetic 
contribution to the vorticity vanishes; this generalizes the results 
in Ref.~\cite{HG&06} where a particular space-time was considered.
%
%
The particular case of proportional real and imaginary parts leads to 
the case of proportional electric and magnetic fields and corresponds 
to the cases studied in Refs.~\cite{Das79,HG&06}.

Note that the considerations above on the vorticity scalar and the Poynting 
vector are completely general and valid for any stationary axially symmetric 
spacetime. 

\begin{center}
\begin{figure}[h!]
\begin{center}
\begin{tabular}{cc}
{\includegraphics[width=\columnwidth]{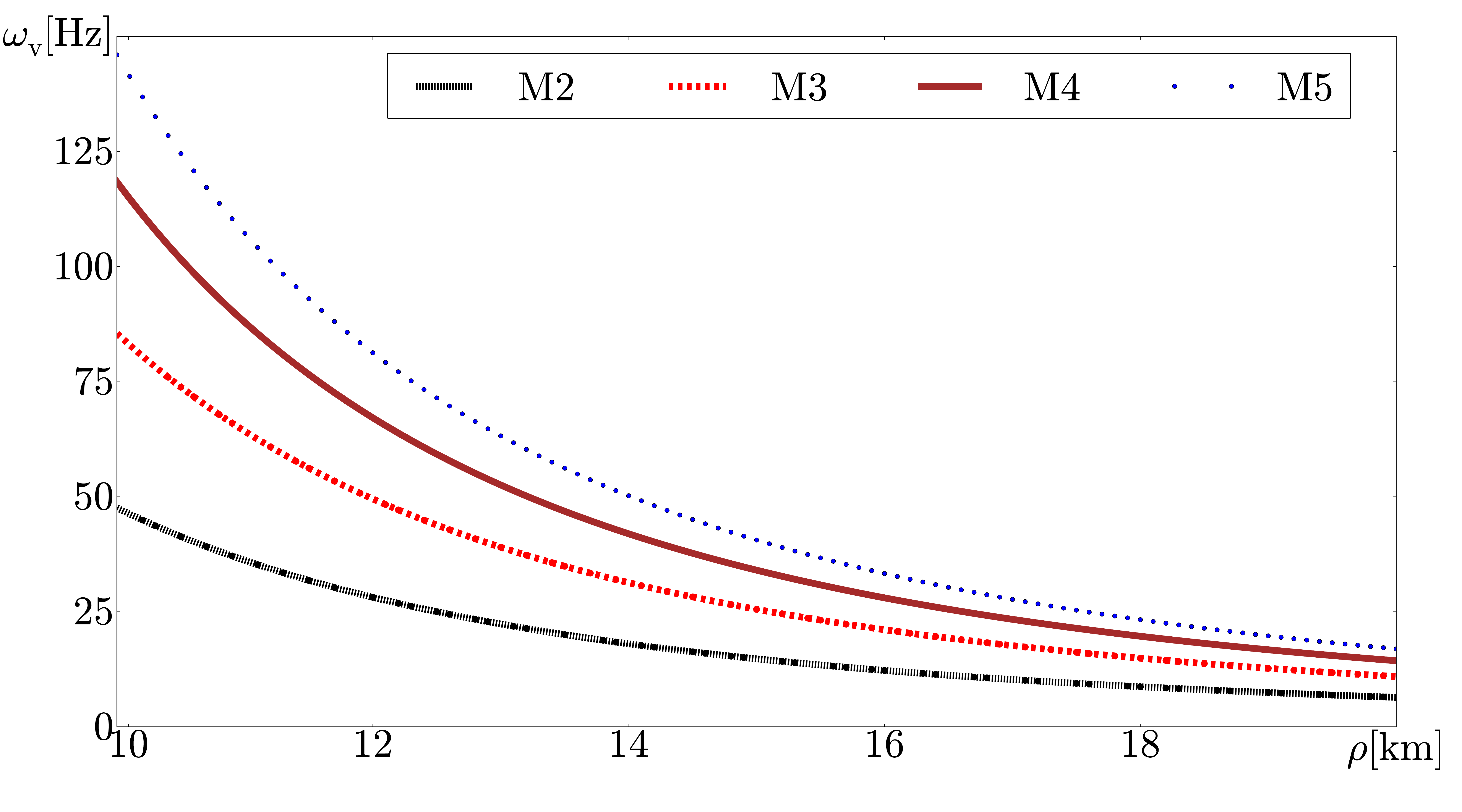}}
\end{tabular}
\end{center}
\caption{Vorticity generated by the mass currents. 
All the electromagnetic moments are zero in this case.
The vorticity of the spacetime was calculated for the models 2 ($j=0.194$),
3 ($j=0.324$), 4 ($j=0.417$) and 5 ($j=0.483$) with the equation of state L 
in Ref.~\cite{Pappas12a}. 
They are listed in the Table~\ref{tp} }
\label{Fig:massvort}
\end{figure}
\end{center}
To study the contribution of the electromagnetic energy flux to the 
vorticity,  characterize first the contributions of the rotation 
of the source. 
In this case, the Ernst potential $\Phi$ that encodes all the electromagnetic 
moments is zero and $\mathcal{E}$ is complex.
For fast rotating stars, the vorticity scalar (and the Lense-Thirring 
effect) is larger than for slow rotating stars, this can be seen in the 
Fig.~(\ref{Fig:massvort}).

When one includes an electromagnetic field to a particular star 
model, the Ernst potential $\Phi$ is no longer zero, and it has an 
important role in the vorticity scalar.
The particular contribution depends on the structure generated 
by the multipole expansion and the Poynting vector.
Before considering the purely electromagnetic contribution to the
frame--dragging in realistic situations (zero or negligible total electric 
charge), consider the case of a source with a dipole magnetic field 
and electric charge.

\begin{center}
\begin{figure}[h!]
\begin{center}
\begin{tabular}{c}
{\includegraphics[width=\columnwidth]{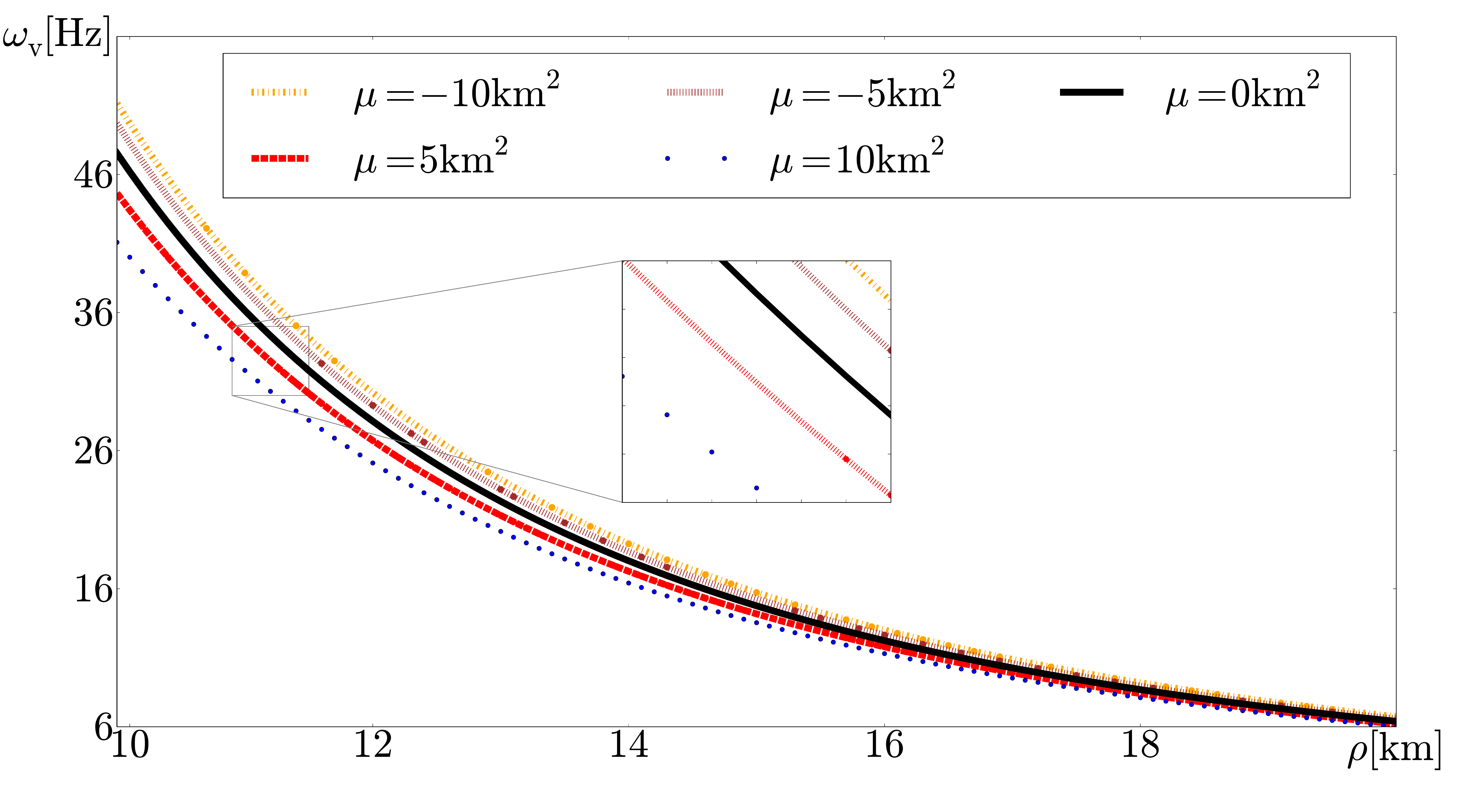}}
\\
{\includegraphics[width=\columnwidth]{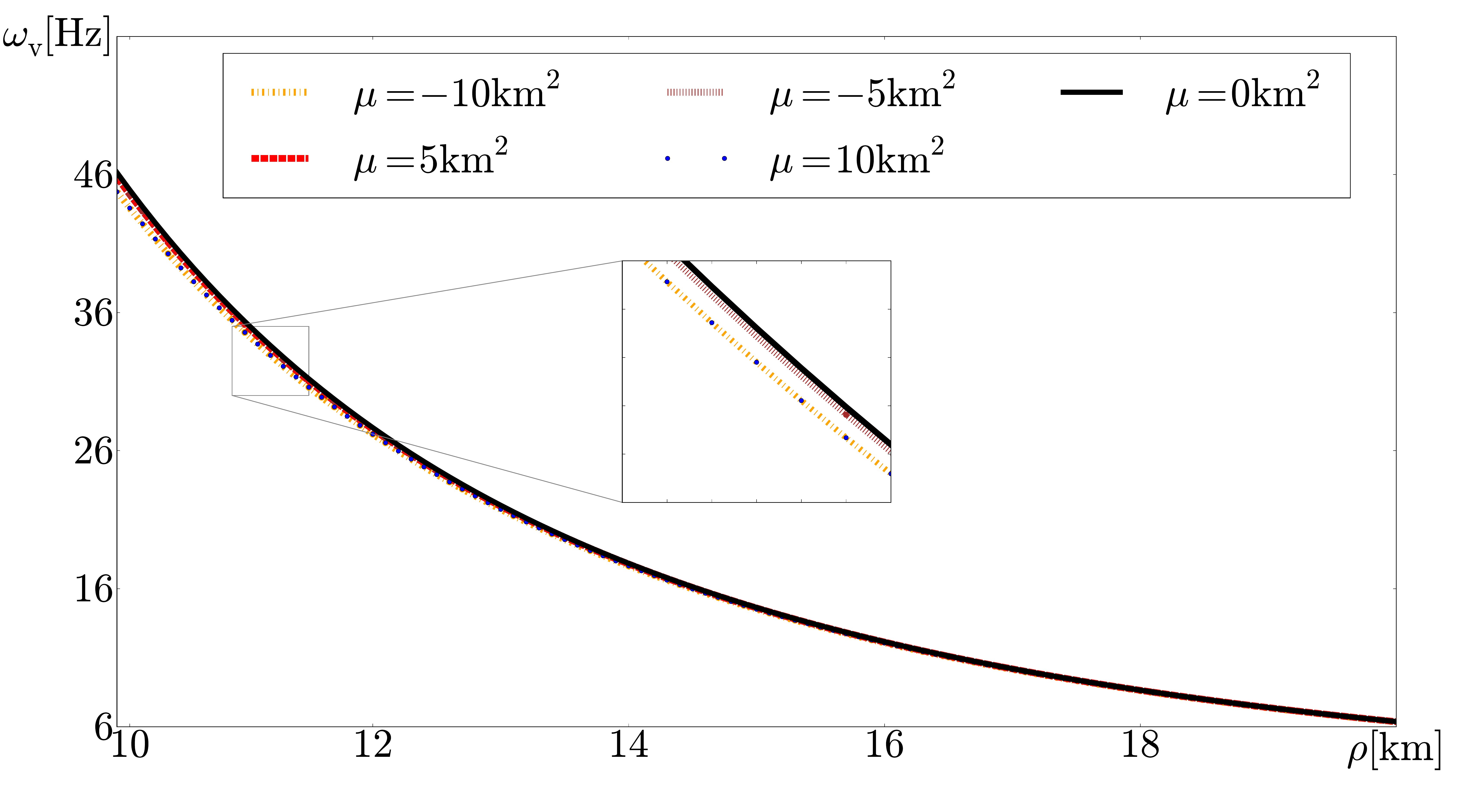}}
\end{tabular}
\end{center}
\caption{Electromagnetically generated vorticity by the presence of a magnetic 
dipole and a fixed electric monopole of $\varsigma=0.1$~km (upper panel with 
$\upsilon=0$, $\chi=0$ and $\zeta=0$)  and an electric dipole of 
$\upsilon=10$~km$^2$ (lower panel with $\varsigma=0$, $\chi=0$ and $\zeta=0$). 
The vorticity of the spacetime was calculated for the model 2  for the equation of 
state L in Ref.~\cite{Pappas12a}. 
Parameters are listed in Table~\ref{tp}. }
\label{Fig:fieldvort1}
\end{figure}
\end{center}

The upper panel of Fig.~(\ref{Fig:fieldvort1}) depicts the functional 
dependence of the vorticity scalar on the distance from the star for a fixed 
value of the electric monopole and for a variety of magnetic dipole moments.
The vorticity scalar decreases (increases) from its value in the vacuum
situation ($\mu=0$) when the dipole is parallel (anti--parallel) to the star's 
rotation--axis.
The reason for this phenomenon can be understood in terms of 
Eq.~(\ref{equ:vorticity}).
To do so, note that $\mathcal{E}$ decreases monotonically 
as a function of $z$ and $\rho$; thus, its derivatives carries 
a negative sign.
Although the same argument applies to $\Phi$, when the sign of $\mu$
changes, see above, the Poynting vector changes its circulation and that
is enough to change the sign of its contribution to the vorticity scalar
[see Eq.~(\ref{equ:vorticity})].
Alternatively, assume that the reflection symmetry exits and set 
the mass currents parameters $a$ and $s$ to zero, for this case, 
$\Phi(\rho,z; a=0, s=0, -\mu) = \Phi^*(\rho,z; a=0, s=0, \mu)$ and 
therefore $\Im \Phi^* \nabla \Phi$ changes sign.

The lower panel of Fig.~\ref{Fig:fieldvort1} depicts the functional dependence 
of the vorticity scalar on the distance from the star for a fixed 
value of the electric dipole and for a variety of magnetic dipole moments.
The contribution in this case is weaker than in the case of an electric
monopole.
However, results in the lower panel of Fig.~\ref{Fig:fieldvort1} are more realistic
than those in the upper panel.
Interestingly, in the lower panel, there is no change in the sign of the 
electromagnetic contribution to the vorticity scalar when $\mu$ changes
sign. 
The reason for this relies on the fact that for $\upsilon\neq0$ or 
$\zeta\neq0$, the reflection symmetry around the equatorial plane breaks down 
and the arguments provided above do not apply, i.e., 
$\Phi(\rho,z; a=0, s=0, -\mu)$ does not equate to 
$\Phi^*(\rho,z; a=0, s=0, \mu)$.

Based on the processes described above on the generation of a Poynting 
vector from the fields induced by the mass currents ($a\neq 0$ and $s\neq0$),
it is clear that even if the parameters associated with the electric field
($\varsigma$, $\upsilon$ and $\chi$) are set to zero, but the parameters 
associated with the magnetic field ($\mu$ and $\zeta$) are non-vanishing,
then a rotationally-induced electromagnetic contribution to the vorticity 
scalar is present.
A similar scenario takes place if the parameters associated with the magnetic
field are set to zero and non-vanishing electric parameters are considered. 
In particular, even in the low rotation regime, a non-negligible electric field is 
generated \cite{Rezzolla21042001} and may be important to characterize the evolution of the 
electromagnetic structure in neutron stars. 
Moreover, in the case of fast rotation, the frame-dragging caused by a Kerr black 
hole significantly distorts the structure of an external magnetic field and this scenario 
may be relevant for low accreting black holes as the one present in the Milky Way 
center \cite{karas2012,karas2014}.

The most relevant contributions to the vorticity scalar from the electromagnetic
field were considered above.
However, in general, all the multipole moments contribute to the vorticity 
scalar and the details of the net result may deviate from those discussed 
above; albeit, the magnitude of the effect is not expected to differ from 
the predictions in Fig.~\ref{Fig:fieldvort1}.

\section{Orbital equatorial Motion and epicyclic frequencies}
In general relativity, as in Newtonian gravitation, all the physical 
characteristics of the source  have an effect on the dynamics of orbiting
of particles. 
One of the most distinctive points between both theories is the 
frame-dragging, but is not an observable by itself.
A way to measure it, is to appeal to the dynamics around the source 
and characterizing the behavior of, e.g., neutral test 
particles~\cite{1998ApJ...492L..59S,GPV13}. 
%

Observationally, the Keplerian motion of matter could be useful to model
the quasi-periodic oscillations (QPOs) that are present in the low-mass 
X-ray binaries (LMXBs) containing a neutron star~\cite{vanderKlis20062675}. 
These oscillations occur at frequencies in the range of $\mathrm{kHz}$ 
and come in pairs, the upper and the lower mode correspond to the 
frequencies of Keplerian motion and periastron precession of the 
accreted matter in the close vicinity of the star \cite{1999ApJ...513..827M}. 
An additional effect is that these equatorial orbits will exhibit a 
relativistic nodal precession due to frame dragging \cite{1998ApJ...492L..59S}, 
causing a detectable signal in the spectra of LMXBs \cite{2041-8205-760-2-L30}. 

The following subsections are devoted to quantifying the magnetic 
quadrupole effect in the ISCO and the epicyclic frequencies for different
models of neutron stars \cite{Pappas12a}.

\subsection*{Influence of the field on the ISCO radii}
The dynamics of an orbiting  particle can be analyzed by using the 
Lagrangian formalism.
To do so, consider a particle of rest mass $\mathfrak{m}_0=1$ 
moving in the space-time described by the metric functions in 
Eq.~(\ref{Papapetrou}), the Lagrangian of the particle is then 
given by
\begin{equation}
\label{lag}
{\cal L}=\frac{1}{2}g_{\mu\nu}\dot{x}^\mu\dot{x}^\nu,
\end{equation}
where the dot denotes differentiation with respect to the proper time 
$\tau$, $x^\mu(\tau)$ are the Weyl-Lewis-Papapetrou coordinates. 
For a stationary and axisymmetric spacetime, there are two constants 
of motion related to the time coordinate $t$  and azimuthal coordinate 
$\phi$. 
Thus, the energy and the canonical angular momentum, respectively, are 
conserved (see Ref.~\cite{1995PhRvD..52.5707R} for details).
Assuming that the motion takes place in the equatorial plane of the 
star $z=0$, and taking into account the four velocity normalization 
for massive particles $g_{\mu\nu}u^\mu u^\nu=-1$, one can identify an 
effective potential that characterizes the motion in the plane (see, 
e.g., Ref. \cite{1972ApJ...178..347B})
\begin{equation}
\label{(9)}
 V_{\mathrm{eff}}(\rho)=1-\frac{E^2g_{\phi \phi}+2ELg_{t\phi}
 +L^2 g_{tt}}{g^2_{t\phi}-g_{\phi \phi}g_{tt}}.
\end{equation}
For a particle moving in a circular orbit, the energy $E$ and 
the canonical angular momentum $L$ are determined by the conditions 
$V_{\mathrm{\mathrm{eff}}}(\rho)=0$ and 
$\mathrm{d}V_{\mathrm{eff}}/\mathrm{d}\rho=0 $ (see, e.g., Ref.~\cite{2002MNRAS.336..831S}). 
The marginal stability condition reads
$\mathrm{d}^2V_{\mathrm{eff}}/\mathrm{d}\rho^2=0$.
Thus, the ISCO's radius is determined by solving numerically the previous 
equation for $\rho$.

The importance of determining the ISCO is that accretion disks extend 
from the last stable orbit to exterior zones, so the ISCO is an inner 
boundary for the accreted matter. 
As stated above, the inclusion of a magnetic field causes a dependence
of all physical quantities of the intensity on the field. 
From the International System of Units (SI), the conversion to 
geometrized units is given by
\begin{equation}
 \mu_{\mathrm{geom}}=\frac{10^{-6}\sqrt{G \mu_{0}}}{c^{2}}\mu_{\mathrm{SI}},
\end{equation}
where $G$ is the gravitational constant, $\mu_{0}$ is the vacuum 
permittivity, $c$ is the speed of light and $\mu$ is given in units of 
Am$^2$.
The units of  $\zeta$ are Am$^{3}$ and the conversion for 
the quadrupole reads
\begin{equation}
\zeta_{\mathrm{geom}}=\frac{10^{-9}\sqrt{G \mu_{0}}}{c^{2}}\zeta_{\mathrm{SI}}.
\end{equation}
In the same way, the electric multipole moments can be written in 
geometrized units as:
\begin{align}
 \varsigma_{\mathrm{geom}}&=\sqrt{\frac{G\mu_0}{4\pi}} \varsigma_{\mathrm{SI}},
 \\
 \upsilon_{\mathrm{geom}}&=10^{-3}\sqrt{\frac{G\mu_0}{4\pi}} \upsilon_{\mathrm{SI}},
 \\
 \chi_{\mathrm{geom}}&=10^{-6} \sqrt{\frac{G\mu_0}{4\pi }} \varsigma_{\mathrm{SI}}.
\end{align}

\begin{figure}[htp!]
     \begin{center}
            \label{2to4}
            \includegraphics[width=\columnwidth]{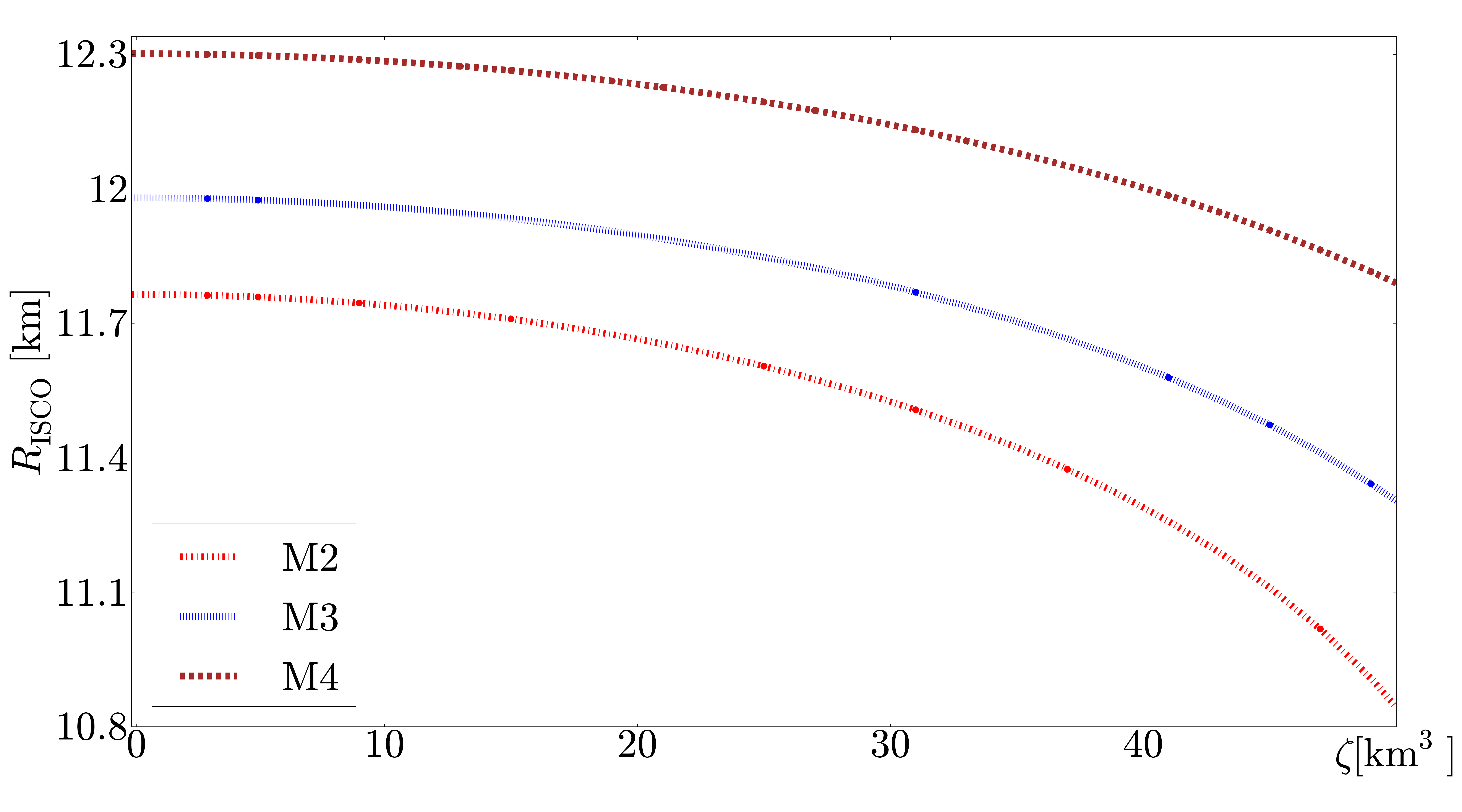}
    \end{center}
    \caption{%
        ISCO radius as a function of magnetic quadrupole parameter $\zeta$, 
        with $\mu=0~\mathrm{km}^2.$ 
	     The physical parameters for the star correspond to the models  
	     2-4  listed in Table~\ref{tp}.
	     We can se that the ISCO radius decreases for increasing $\zeta$.
     }%
   \label{isco}
\end{figure}

Consider a superposition of a fixed dipolar and  quadrupole component 
that will be varied from $1$ km$^{3}$ to $50$ km$^{3}$. 
Figure~\ref{isco} depicts the ISCO radius as a function of the parameter $\mu$
and $\zeta$ for three realistic numerical solutions for rotating 
neutron stars models derived in Ref.~\cite{Pappas12a}. 
The parameter $\mu$ is set to $1$ km$^2$ that corresponds to a magnetic 
dipole field of $10^{12}$~T.
The quadrupole parameter $\zeta$ is chosen between $0$  and $50$~km$^3$ 
that corresponds to magnetic quadrupoles from $0$ to $5\times10^{35}$~Am$^3$, 
respectively.
In all these cases, the ISCO radius decreases for increasing $\zeta$. 
This can be explained as a result of the deformation of the space-time 
by the energy stored in the electromagnetic (see, e.g., Ref.~\cite{GPV13}).
We elaborate more on this in the next section.

\subsection*{Influence of the electromagnetic field in the Keplerian 
and the epicyclic frequencies}
\begin{figure}[htp!]
     \begin{center}
            \label{lense2to4}
            \includegraphics[width=\columnwidth]{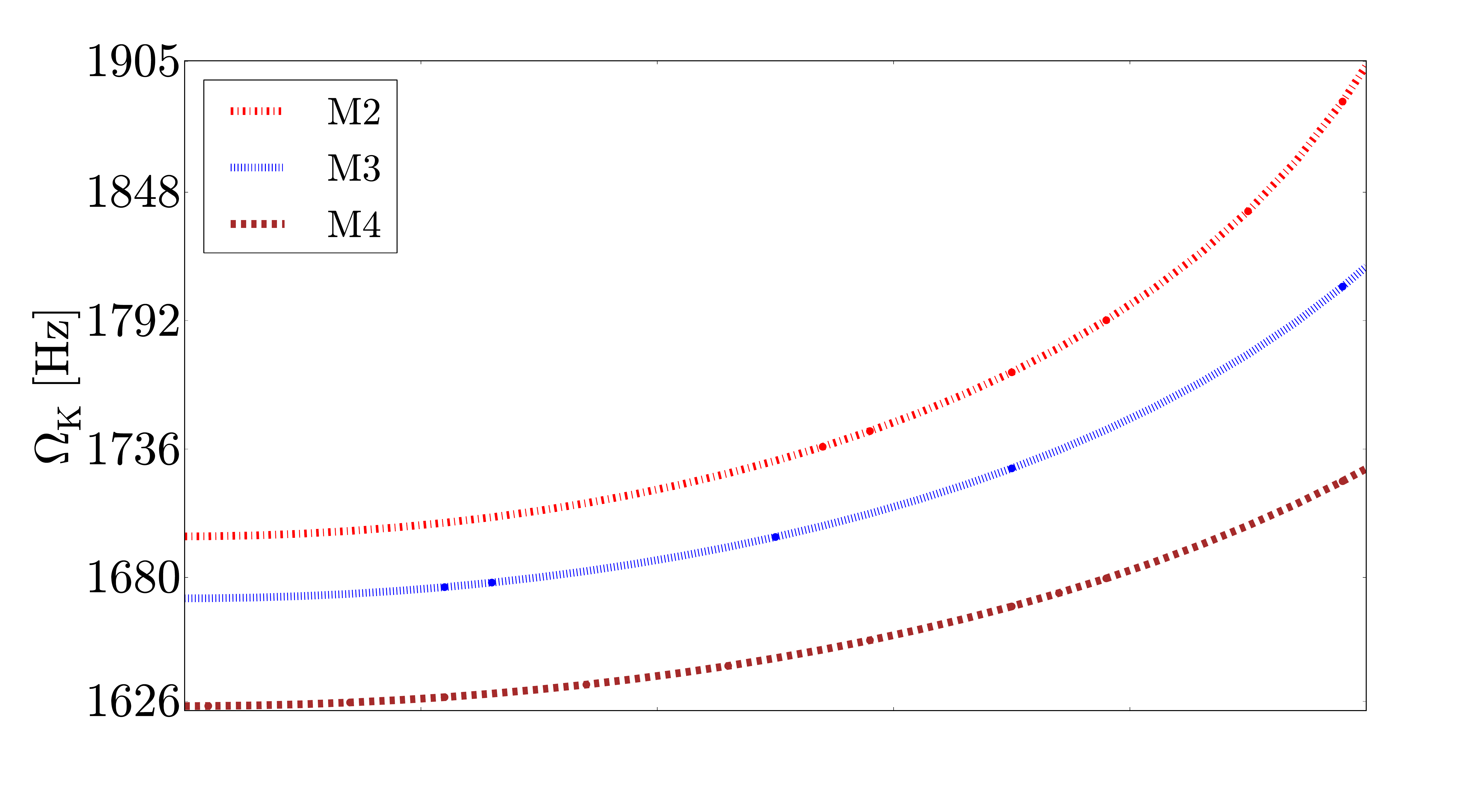}
       \vspace{-1.0cm} \\%
           \label{fig:second}
           \includegraphics[width=\columnwidth]{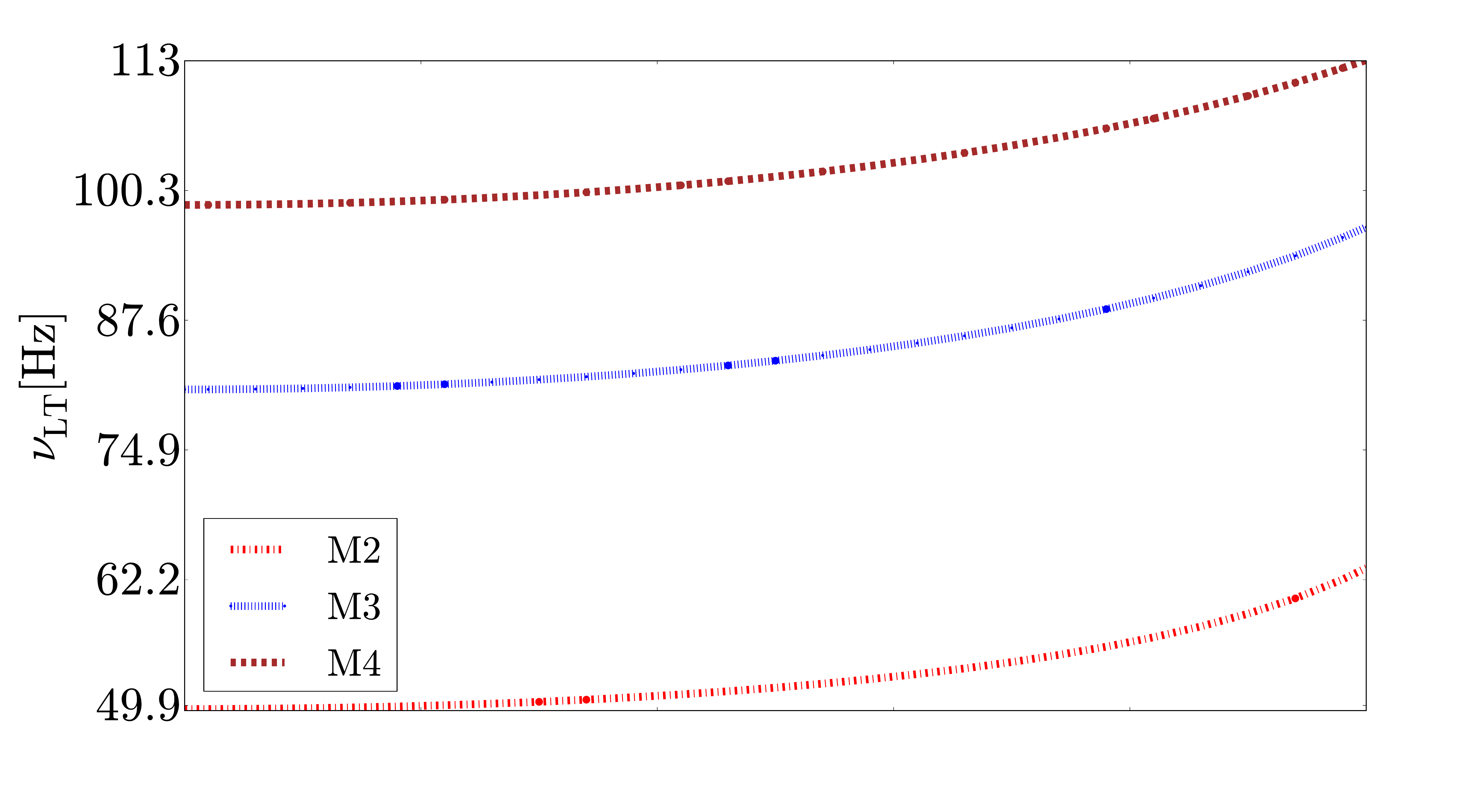}
        \vspace{-1.0cm} \\ 
            \label{fig:third}
            \includegraphics[width=\columnwidth]{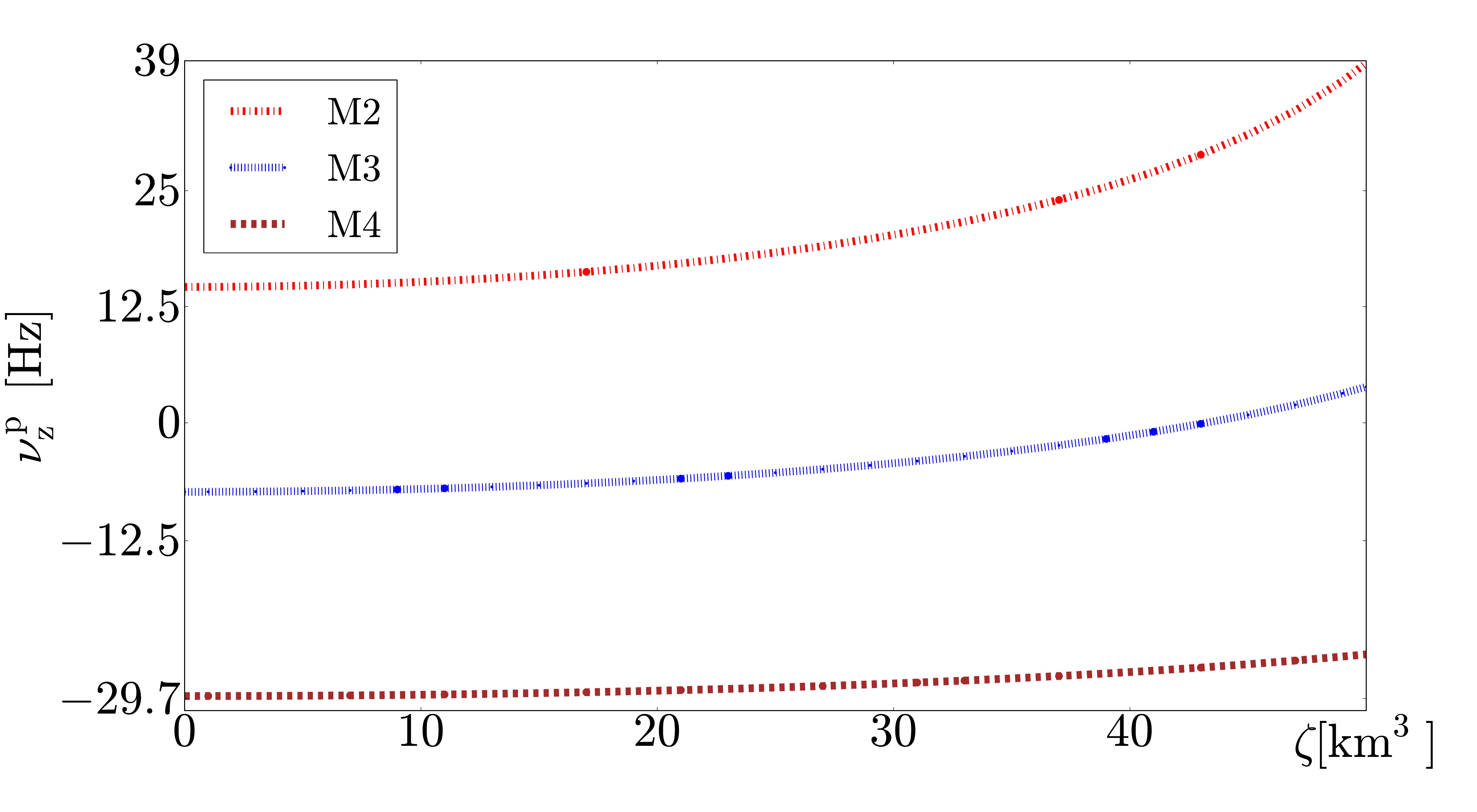}        
            \vspace{-1.0cm}
    \end{center}
    \caption{%
        Influence of the magnetic quadrupole moment $\zeta$ in the 
        epicyclic frequencies for different models of neutron stars
        with $\varsigma=0$, $\mu=1$~km$^2$, $\upsilon=0$ and $\chi=0$.
        Vacuum parameters of the star are listed in Table~\ref{tp}.
     }%
   \label{fig:epfrec}
\end{figure}

Imposing the conditions of constant orbital radius, 
$\mathrm{d}\rho/\mathrm{d}\tau=0$ and taking into account that
$\mathrm{d}\phi/\mathrm{d}\tau=\Omega_\mathrm{K} \mathrm{d}t/\mathrm{d}\tau$, 
the Keplerian frequency reads (see, e.g., Ref.~\cite{1995PhRvD..52.5707R})
\begin{equation}
\label{8}
 \Omega_{\mathrm{K}}=\frac{\mathrm{d}\phi}{\mathrm{d}t}=
 \frac{-g_{t \phi,\rho} \pm \sqrt{(g_{t\phi,\rho})^2
 -g_{\phi \phi,\rho}g_{tt,\rho}}}{g_{\phi \phi,\rho}},  
\end{equation}
where ``$+$'' and ``$-$''  denotes the Keplerian frequency for 
corotating and counter-rotating orbits, respectively.
In Ref.~\cite{GPV13} the functional dependence of the Keplerian 
frequency on the magnetic dipole parameter $\mu$ was discussed, 
here interest is in the functional dependence on the magnetic
quadupole parameter $\zeta$.
The upper panel of Fig.~\ref{fig:epfrec} depicts the functional dependence
on  $\zeta$, as in the case analyzed in Ref.~\cite{GPV13}, the Keplerian
frequency increases with increasing $\zeta$ because the ISCO's radius decreases 
(see Fig.~\ref{isco}) as a consequence of the additional deformation of the
space-time by the electromagnetic field.
In particular, the energy of the electromagnetic field shifts the marginally
unstable region toward the star.

Analytical expressions for the radial and vertical frequencies follow from 
allowing slightly radial and vertical perturbations of the orbit.
According to Ref.~\cite{1995PhRvD..52.5707R}, the radial and vertical 
epicyclic frequencies are given by
\begin{align}
\label{9}
 \nu_{\alpha}= \frac{1}{2\pi}\bigg\{-\frac{g^{\alpha\alpha}}{2}\bigg[(g_{tt}
 +g_{t\phi}\Omega_{k})^2\bigg(\frac{g_{\phi\phi}}{\rho^2}\bigg)_{,\alpha\alpha} 
\nonumber \\
 - 2(g_{tt}+g_{t\phi}\Omega_{k})(g_{t\phi}+g_{\phi\phi}\Omega_{k})
 \bigg(\frac{g_{t \phi}}{\rho^2}\bigg)_{,\alpha\alpha}
\\ \nonumber
+
(g_{t\phi}+g_{\phi\phi}\Omega_{k})^2\bigg(\frac{g_{tt}}{\rho^2}
\bigg)_{,\alpha\alpha}\bigg]\bigg \},
\end{align}
with $\alpha=\{\rho,z\}$.
In the relativistic precession model (RPM) \citep{Stella99}, 
the periastron $\nu_\rho^{\mathrm p}$ and the nodal $\nu_z^{\mathrm p}$ 
frequencies are the observational relevant ones, they are defined by
\begin{equation}
 \nu^{\mathrm p}_{\alpha}=\frac{\Omega_{\mathrm{K}}}{2\pi}-\nu_{\alpha}.
\end{equation}
At the ISCO, the radial oscillation frequency equals the Keplerian 
frequency and only the vertical precession $\alpha=z$ is considered 
here.
The influence of the magnetic quadrupole on the vertical precession 
frequency is depicted in the lower panel of Fig.~\ref{fig:epfrec}.
The behaviour and the underlying physical mechanism are analogous 
to those for the Keplerian frequency.

The frequency $\nu_{\mathrm{LT}}$ that characterizes the Lense--Thirring 
effect is given by (see, e.g.,  Ref.~\cite{1995PhRvD..52.5707R})
\begin{equation}
\nu_{\mathrm{LT}}=-\frac{1}{2\pi}\frac{g_{t\phi}}{g_{\phi\phi}}\ ,
\end{equation}
and could be relevant to model the  horizontal branch oscillations observed 
in the Low Mass X-Rax Binaries(LMXBs) \citep{Stella99}.
The strength of the frame--dragging effect increases for fast rotating 
objects and, as shown in the central panel of Fig.~\ref{fig:epfrec}, 
for stronger magnetic fields.
Figure~\ref{fig:epfrec} also shows that the electromagnetically-induced frame 
dragging has a direct effect on the orbiting particles, even if they are neutral.

\section{Concluding remarks}
We present a new stationary axisymmetric nine-parameter closed-form
analytic solution that generalizes the Kerr solution with arbitrary
mass-quadrupole moment, octupole current moment, electric and magnetic 
dipole and electric and magnetic quadrupole moments.
The analytic form of its multipolar structure and their electric and 
magnetic fields are also presented. 
According to the arguments presented through the paper, this solution could 
be used to model the exterior gravitational and electromagnetic fields around 
strongly magnetized stars, in particular white dwarfs.  
Besides, this model also could be used even for the description of exotic stars 
such as the $\tau$ Sco recently reported by Donati \textit{et al.} \cite{Donati06}.

This solution allowed for a comprehensive analysis of the contribution of complex 
and intense electromagnetic fields to quintessential hallmark of general relativity, 
namely, the Lense--Thirring effect.
It was shown that if the value of the parameters is such that the reflection
symmetry is preserved \cite{PS06}, then a sign change of the magnetic dipole 
parameter $\mu$ is enough to change the direction of the circulation of the 
Poynting vector and the sign of the electromagnetic contribution to the vorticity
scalar.
The influence of complex electromagnetic fields on the ISCO's radius and the
epicyclic frequencies were also considered and it was shown that for strong 
magnetic fields, the influence is not negligible.

Additional interest is in studying the effect of the magnetic field 
in the quadrupole moment of the star (see, e.g., 
Refs.~\cite{Mastrano11092013,HC&13}). 
This topic has gained recently theoretical and observational interest and 
is currently discussed under the title of $I-Love-Q$ relations \cite{HC&13}.
Due to the great generality of the present analytic exact solution, this subject 
will be explored in a forthcoming contribution \cite{GVP15}.

\section*{Acknowledgments}
Fruitful discussions with Prof. C\'esar A. Valenzuela-Toledo are acknowledged with 
pleasure.
This work was partially supported by Fundaci\'on para la Promoci\'on de
la Investigaci\'on y la Tecnolog\'ia del Banco la Rep\'ublica
Grant No. 2879. and by Comit\'e para el Desarrollo de la
Investigaci\'on (CODI) of Universidad de Antioquia under
the Estrategia de Sostenibilidad 2015-2016.

\bibliography{NonDipFieldsv2}

\appendix

\begin{widetext}
\section{Metric Functions of the Developed Solution}
\label{app:metricfuncs}%
This appendix summarizes the relevant equations of the developed metric. 
The potentials in the symmetry axis can be written as \cite{RMM95}:
\begin{equation}
e(z)=1+\displaystyle \sum_{i=3}^{3} \frac{e_{i}}{z-\beta_{i}}\,
,\qquad f(z)=\displaystyle \sum_{i=3}^{3}
\frac{f_{i}}{z-\beta_{i}}\, ,
\end{equation}
with
\begin{eqnarray}
    e_{j}= (-1)^{j}\frac{2 m
    \beta^{2}_{j}}{(\beta_{j}-\beta_{k})(\beta_{j}-\beta_{i})}\, ,
\qquad
    f_{j}= \frac{i \zeta +(\varsigma + i \mu )\beta_{j}}
    {(\beta_{j}-\beta_{k})(\beta_{j}-\beta_{i})}\, , \quad i,k \neq
    j\, .
\end{eqnarray}

Then,  the Ernst potentials and the metric functions in whole spacetime 
are derived with the aid of the Sibgatullin's integral method \cite{Sib91,MS93}. 
By using the representation proposed in Ref.~\cite{Breton99} and used also Ref.~\cite{PRS06}, 
\begin{equation}
{\cal E}=\frac{A + B }{A - B}\, , \qquad \Phi=\frac{C}{A - B}\, ,
\label{ernstpot}
\end{equation}
\begin{align*}
f&=\frac{A \bar{A}-B \bar{B} + C \bar{C}}{( A -
B)(\bar{A}-\bar{B})}, 
\quad 
e^{2\gamma}=\frac{A \bar{A} -B \bar{B} + C \bar{C}}{
\displaystyle{K \bar{K}\prod_{n=1}^{6}r_n}},
\quad 
\omega = \frac{{\rm Im}[(A + B)\bar{H}-(\bar{A} + \bar{B})G - C
\bar{I}]}{A \bar{A} - B \bar{B} + C \bar{C}}\, ,
\end{align*}
where
\begin{align*}
    A &= \displaystyle\sum_{{\small 1\leq i < j < k \leq 6}} a_{i j\,
    k} r_{i}\,r_{j}\,r_{k}\, ,
\quad
    B =\displaystyle\sum_{{\small
    1\leq i < j \leq 6}} b_{i j} r_{i}\,r_{j},
\quad
    C = \displaystyle\sum_{{\small 1\leq i < j \leq 6}} c_{i j}
    r_{i}\,r_{j}\, ,
\quad
    K = \displaystyle\sum_{{ \small 1\leq i < j < k \leq 6}} a_{i j\,
    k}\, ,    
\\     
    H &= z\,A - (\beta_{1} + \beta_{2}+
    \beta_{3})B
    +\displaystyle \sum_{{\small 1\leq i < j < k \leq 6}} h_{i j\, k} r_{i}\,r_{j}\,r_{k} +
    \displaystyle\sum_{{\small 1\leq i < j \leq 6}} (\alpha_{i} + \alpha_{j})\,b_{i j}
    \,r_{i}\,r_{j},
\\
    G &= -(\beta_{1} + \beta_{2} + \beta_{3})\,A + z\,B + \displaystyle\sum_{{\small 1\leq i < j \leq
    6}} g_{i j} \,r_{i}\,r_{j}+ \displaystyle \sum_{{\small 1\leq i < j < k \leq 6}} (\alpha_{i} +
    \alpha_{j} + \alpha_{k})a_{i j\,k} r_{i}\,r_{j}\,r_{k},
\\
    I &= (f_{1} + f_{2} + f_{3})(A - B) + (\beta_{1}+\beta_{2} +
    \beta_{3} - z)\, C
    + \displaystyle\sum_{{\small 1\leq i < j < k \leq 6}} p_{i j\,k}
    r_{i}\,r_{j}\,r_{k} + \displaystyle\sum_{{\small i=1}}^{6} p_{i}\,
    r_{i}
\\
    &+ \displaystyle\sum_{{\small 1\leq i < j \leq 6}} [p_{i
    j}-(\alpha_{i} + \alpha_{j})c_{ij}] r_{i}\,r_{j},
\end{align*}
with
\begin{align*}
r_i &= \sqrt{\rho^2 + (z-\alpha_i)^2}\, , 
\quad
a_{i j\,k} = (-1)^{i + j + 1}\Lambda_{i j k}\,\Gamma_{l | m n}\, ,
\quad
b_{i j} = (-1)^{i + j}\lambda_{i j}\,H_{l | m n p}\, ,
\\
c_{i j} &= (-1)^{i + j}\lambda_{i j}[f(\alpha_l)\,\Gamma_{m | n p}
- f(\alpha_m)\,\Gamma_{n | p l}+ f(\alpha_n)\,\Gamma_{p | l m}
- f(\alpha_p)\,\Gamma_{l | m n}]\, ,
\\
h_{i j\,k} &= (-1)^{i + j + k}\Lambda_{i j k}(e^{*}_1\,\delta_{2 3
| l m n} + e^{*}_2\,\delta_{3 1 | l m n}+ e^{*}_3\,\delta_{1 2
| l m n})\, ,
\\
g_{i j} &= (-1)^{i + j}\lambda_{i j}(\alpha_{l}\,\Gamma_{m | n p} -
\alpha_{m}\,\Gamma_{n | p l}+ \alpha_{n}\,\Gamma_{p | l m} -
\alpha_{p}\,\Gamma_{l | m n})\, ,
\\
p_i &= (-1)^i D_{i}[f(\alpha_l)\,H_{m | n p s} - f(\alpha_m)\,H_{n
| p s l}+ f(\alpha_n)\,H_{p | s l m} - f(\alpha_p)\,H_{s | l m
n}+ f(\alpha_s)\,H_{l | m n p}]\, ,
\\
p_{i j} &= (-1)^{i + j} \lambda_{i j}(e^{*}_1\,\Upsilon_{2 3 | l m
n p} + e^{*}_2\,\Upsilon_{3 1 | l m n p}+ e^{*}_3\,\Upsilon_{1
2 | l m n p})\, ,
\\
p_{i j\,k} &= (-1)^{i + j +1} \Lambda_{i j\,k}(e^{*}_1\,\Psi_{2 3 |
l m n} + e^{*}_2\,\Psi_{3 1 | l m n}+ e^{*}_3\,\Psi_{1 2 | l m
n})\, ,
\end{align*}
and
\begin{align*}
\lambda_{i j} &= (\alpha_i - \alpha_j)\,D_i\,D_j\, ,
\quad
\Lambda_{i j\,k} = (\alpha_i - \alpha_j)(\alpha_i -
\alpha_k)(\alpha_j - \alpha_k)\,D_i\,D_j\,D_k\, ,
\quad
D_i = \frac{1}{(\alpha_i - \beta_1)(\alpha_i - \beta_2)(\alpha_i -
\beta_3)}\, ,
\\
\Gamma_{l | m n} &= H_3(\alpha_l)\,\Delta_{1 2 | m n} +
H_3(\alpha_m)\,\Delta_{1 2 | n l} + H_3(\alpha_n)\,\Delta_{1 2
| l m}\, ,
\quad
\Delta_{l m | n p} = H_{l}(\alpha_n)\,H_{m}(\alpha_p) -
H_{l}(\alpha_p)\,H_{m}(\alpha_n)\, ,
\\
H_{l}(\alpha_n) &= \frac{2 \prod_{p \neq n} (\alpha_p -
\beta^{*}_l)}{\prod_{k \neq l}^{3} (\beta^{*}_l -
\beta^{*}_k)\,\prod_{k = 1}^{3} (\beta^{*}_l - \beta_k)}
- 2\,\sum_{k = 1}^{3} \frac{f^{*}_l\,f_k}{(\beta^{*}_l -
\beta_k)(\alpha_n - \beta_k)}\, ,
\quad
\delta_{l m | n p s} = \Delta_{l m | n p} + \Delta_{l m | p s} + \Delta_{l m | s n},
\\
h_{l | m n p} &= H_3(\alpha_l)\,\delta_{1 2 | m n p}\, ,
\quad
H_{l | m n p} = h_{l | m n p} + h_{m | n p l} + h_{n | p l m} +
h_{p | l m n}\, ,\\
\Psi_{l m | n p s} &= f(\alpha_n)\,\Delta_{l m | p s} +
f(\alpha_p)\,\Delta_{l m | s n} + f(\alpha_s)\,\Delta_{l m | n p}\, ,
\\
\Upsilon_{l m | n p r s} &= f(\alpha_n)\,\delta_{l m | p r s} -
f(\alpha_p)\,\delta_{l m | r s n}+f(\alpha_r)\,\delta_{l m | s n
p} - f(\alpha_s)\,\delta_{l m | n p r}\, ,
\end{align*}
being $\alpha$'s the roots of the Sibgatullin's equation
\cite{Sib91,MS93}
\begin{equation}\label{eq sibga}
e(z)+\tilde{e}(z)+2{\tilde{f}}(z)f(z)=0.
\end{equation}

\section{Electromagnetic Field: Analytic Form}
\label{app:emfield}%
The $A_t$ potential is the real part of the electromagnetic Ernst
potential $\Phi$, and the
potential $A_{\phi}$ can be calculated as the real part of the
Kinnersley potential ${\cal K}= A_{\phi}+iA'_{t}$ \cite{kinn77},
which can be obtained using the Sibgatullin's method and can be
written as
\begin{equation}\label{kinnersleypotential} {\cal
K}=-i\,\frac{I\,(f_{1} + f_{2})}{A - B}\, .
\end{equation}
Thus, the closed--form expressions for the electric and magnetic
fields are
\begin{align*}
    E_\rho &=-\frac{\Lambda}{\sqrt{|A|^2 -
    |B|^2 + |C|^2}}\,{\rm Re}\Bigg\{ \left[ \frac{ C_{,\rho} -
    C\ln (A - B)_{,\rho}}{A - B} \right] \Bigg\}\, ,
\\
    E_z &=-\frac{\Lambda}{\sqrt{|A|^2 - |B|^2 +
    |C|^2}}\,{\rm Re}\Bigg\{ \left[ \frac{ C_{,z} - C\ln (A - B)_{,z}}{A
    - B} \right] \Bigg\}\, ,\nonumber\
\end{align*}
\begin{align*}
    B_\rho&= \frac{{\rm Im} (A - B)\bar{H} +(\bar{A} - \bar{B})G -
    C\bar{I}]}{\rho |A - B|^2}E_z \label{B}+
    \frac{\Lambda \sqrt{|A|^2 - |B|^2 +
    |C|^2}}{\rho |A - B|^2}
    {\rm Im}\Bigg\{\frac{(\bar{f_1} + \bar{f_2})\left[ \bar{I}_{,z}
    -\bar{I}\ln (\bar{A} - \bar{B})_{,z}\right] }{A-B} \Bigg\},
\\
    B_z&=- \frac{{\rm Im} (A - B)\bar{H} +(\bar{A} - \bar{B})G -
    C\bar{I}]}{\rho |A - B|^2}E_{\rho}+ \frac{\Lambda\sqrt{|A|^2 - |B|^2
    + |C|^2}}{\rho |A - B|^2}{\rm Im}
    \Bigg\{\frac{(\bar{f_1} + \bar{f_2})\left[ \bar{I}_{,\rho}
    -\bar{I}\ln (\bar{A} - \bar{B})_{,\rho}\right] }{A-B}
    \Bigg\}.
\end{align*}
when $$\Lambda=\frac{|K|^2\,\prod_{n=1}^{6}r_n}{|A - B|}.$$ 
\end{widetext}

\end{document}